\documentclass[fleqn,usenatbib,useAMS]{mnras}

\usepackage{newtxtext,newtxmath}

\usepackage[T1]{fontenc}

\DeclareRobustCommand{\VAN}[3]{#2}
\let\VANthebibliography\thebibliography
\def\thebibliography{\DeclareRobustCommand{\VAN}[3]{##3}\VANthebibliography}

\usepackage{graphicx}	
\usepackage{amsmath}	
\usepackage{xspace}	
\usepackage{color}
\usepackage{lineno}
\usepackage{tablefootnote}
\usepackage{fleqn,natbib}
\usepackage{flushend}
\usepackage{cuted}
\usepackage{threeparttable}
\usepackage{xcolor}
\usepackage{dcolumn}
\usepackage{multicol}
\usepackage{adjustbox}
\usepackage{float}
\usepackage{multirow}
\usepackage{caption}
\usepackage{supertabular}
\usepackage{subcaption}
\usepackage{longtable}
\usepackage[utf8]{inputenc}

\newcommand{\corr}{}

\newcommand{\sw}[1]{\texttt{#1}}
\newcommand{\af}{\sw{afterglowpy}}
\newcommand{\sxt}{\sw{SExtractor}}
\newcommand{\swp}{\sw{SWarp}}
\newcommand{\solf}{\sw{solve-field}}
\newcommand{\psfex}{\sw{PSFEx}}

\newcommand{\thisgrb}{GRB~210204A\xspace}

\newcommand{\asat}{\corr{\textit{AstroSat}}}
\newcommand{\fermi}{\corr{\textit{Fermi}}}
\newcommand{\fermigbm}{\corr{\textit{Fermi Gamma-Ray Burst Monitor}}}
\newcommand{\integral}{\corr{\textit{Integral}}}
\newcommand{\kw}{\corr{\textit{Konus}-Wind}}
\newcommand{\bal}{\corr{\textit{ BALROG}}}
\newcommand{\ztf}{\corr{\textit{ Zwicky Transient Facility}}}
\newcommand{\neil}{\corr{\textit{ Neil Gehrels Swift Observatory}}}
\newcommand{\doot}{\corr{\textit{ Devasthal Optical Telescope}}}
\newcommand{\dfot}{\corr{\textit{ Devasthal Fast Optical Telescope}}}
\newcommand{\git}{\corr{\textit{ GROWTH-India Telescope}}}
\newcommand{\hct}{\corr{\textit{ Himalayan Chandra Telescope}}}

\newcommand{\gmrt}{\corr{\textit{Giant Metrewave Radio Telescope}}}
\newcommand{\bep}{\corr{\textit{ BeppoSAX}}}
\newcommand{\mars}{\corr{\textit{ Mars-Odyssey}}}
\newcommand{\swift}{\corr{\textit{ Swift}}}
\newcommand{\swiftxrt}{\corr{\textit{ Swift X-ray telescope}}}

\newcommand{\fermiT}{{T$_{\rm 0}$}\xspace}
\newcommand{\keV}{{\rm keV}\xspace}
\newcommand{\GHz}{{\rm GHz}\xspace}

\newcommand{\fluence}{\ensuremath{\mathrm{erg~cm}^{-2}}}
\newcommand{\tninty}{{$T_{\rm 90}$}\xspace}

\newcommand{\Ep}{$E_{\rm p}$\xspace}

\newcommand{\iitb}{1}
\newcommand{\lsstc}{2}
\newcommand{\aries}{3}
\newcommand{\ddu}{4}
\newcommand{\maryland}{5}
\newcommand{\jssi}{6}
\newcommand{\iia}{7}
\newcommand{\ncra}{8}
\newcommand{\minnesota}{9}
\newcommand{\berkeley}{10}
\newcommand{\miller}{11}
\newcommand{\stockholm}{11}
\newcommand{\prsu}{13}
\newcommand{\berkleya}{14}
\newcommand{\berkleyb}{15}
\newcommand{\ljmu}{17}
\newcommand{\mini}{18}
\newcommand{\caltech}{19}

\title[Late flaring in \thisgrb]{The long-active afterglow of \thisgrb: Detection of the most delayed flares in a Gamma-Ray Burst}

\author[H. Kumar et al.]{
Harsh Kumar,$^{\iitb,\lsstc}$\thanks{E-mail: harshkumar@iitb.ac.in}
Rahul Gupta,$^{\aries,\ddu}$\thanks{E-mail: rahulbhu.c157@gmail.com}
Divita Saraogi,$^{\iitb}$
Tom{\'a}s Ahumada,$^{\maryland}$
Igor Andreoni,$^{\jssi}$
\newauthor
G.C. Anupama,$^{\iia}$
Amar Aryan,$^{\aries,\ddu}$
Sudhanshu Barway,$^{\iia}$
Varun Bhalerao,$^{\iitb}$
Poonam Chandra,$^{\ncra}$
\newauthor
Michael W. Coughlin,$^{\minnesota}$
Dimple$^{\aries, \ddu}$,
Anirban Dutta$^{\iia}$,
Ankur ghosh$^{\aries}$,
Anna Y. Q. Ho$^{\berkeley,\miller}$,
\newauthor
E. C. Kool,$^{\stockholm}$,
Amit Kumar$^{\aries, \prsu}$,
Michael S. Medford$^{\berkleya, \berkleyb}$,
Kuntal Misra$^{\aries}$,
Shashi B. Pandey$^{\aries}$,
\newauthor
Daniel A. Perley$^{\ljmu}$,
Reed Riddle$^{\caltech}$,
Amit Kumar Ror$^{\aries}$,
Jason~M.~Setiadi$^{\mini}$,
Yuhan Yao$^{\caltech}$
\\
$^{\iitb}$Physics Department, Indian Institute of Technology Bombay, Powai, 400 076, India\\
$^{\lsstc}$LSSTC DSFP Fellow-2018\\
$^{\aries}$Aryabhatta Research Institute of Observational Sciences, Manora Peak, Nainital - 263 001, India\\
$^{\ddu}$Department of Physics, Deen Dayal Upadhyaya Gorakhpur University, Gorakhpur 273009, India\\
$^{\maryland}$Department of Astronomy, University of Maryland, College Park, MD 20742, USA\\
$^{\jssi}$Joint Space Science Institute: College Park, Maryland, US \\
$^{\iia}$Indian Institute of Astrophysics, 2nd Block 100 Feet Rd, Koramangala Bangalore, 560 034, India\\
$^{\ncra}$Swarna Jayanti Fellow, Department of Science \& Technology, India. National Centre for Radio Astrophysics, \\ Tata Institute of Fundamental Research, Ganeshkhind, Pune 411007, India\\
$^{\minnesota}$School of Physics and Astronomy, University of Minnesota, Minneapolis, Minnesota 55455, USA \\
$^{\berkeley}$Department of Astronomy, University of California, Berkeley, 94720, USA\\
$^{\miller}$Miller Institute for Basic Research in Science, 468 Donner Lab, Berkeley, CA 94720, USA\\
$^{\stockholm}$ The Oskar Klein Centre, Department of Astronomy, Stockholm University, AlbaNova, SE-10691, Stockholm, Sweden\\
$^{\prsu}$School of Studies in Physics and Astrophysics, Pandit Ravishankar Shukla University, Raipur, Chattisgarh-492010, India \\
$^{\berkleya}$Department of Astronomy, University of California, Berkeley, Berkeley, CA 94720\\
$^{\berkleyb}$Lawrence Berkeley National Laboratory, 1 Cyclotron Rd., Berkeley, CA 94720\\
$^{\ljmu}$Astrophysics Research Institute, Liverpool John Moores University, IC2, Liverpool Science Park, 146 Brownlow Hill, Liverpool L3 5RF, UK\\
$^{\mini}$University of Minnesota, School of Statistics, 313 Ford Hall, 224 Church Street SE, Minneapolis, MN 55455, USA\\
$^{\caltech}$Division of Physics, Mathematics, and Astronomy, California Institute of Technology, Pasadena, CA 91125, USA\\
}

\date{Accepted XXX. Received YYY; in original form ZZZ}

\pubyear{2022}

\begin{document}
\label{firstpage}
\pagerange{\pageref{firstpage}--\pageref{lastpage}}
\maketitle

\begin{abstract}
We present results from extensive broadband follow-up of \thisgrb over the period of thirty days. We detect optical flares in the afterglow at $7.6\times 10^5$~s and $1.1\times10^6$~s after the burst: the most delayed flaring ever detected in a GRB afterglow. 
At the source redshift of 0.876, the rest-frame delay is $5.8\times10^5$~s (6.71~d). We investigate possible causes for this flaring and conclude that the most likely cause is a refreshed shock in the jet. The prompt emission of the GRB is within the range of typical long bursts: it shows three disjoint emission episodes, which all follow the typical GRB correlations. 
This suggests that \thisgrb might not have any special properties that caused late-time flaring, and the lack of such detections for other afterglows might be resulting from the paucity of late-time observations.
Systematic late-time follow-up of a larger sample of GRBs can shed more light on such afterglow behaviour. Further analysis of the \thisgrb shows that the late time bump in the light curve is highly unlikely due to underlying SNe at redshift (z) = 0.876 and is more likely due to the late time flaring activity. The cause of this variability is not clearly quantifiable due to the lack of multi-band data at late time constraints by bad weather conditions. The flare of \thisgrb is the latest flare detected to date.
\end{abstract}



\section{Introduction} 
\label{sec:intro}
Long Gamma-Ray Bursts (GRBs) originate from the core collapse of massive stars \citep{1993ApJ...413L.101K, 2015PhR...561....1K}. The GRB emission consists of two distinct phases: the prompt emission typically observed in soft $\gamma$-rays and hard X-rays, and the afterglow, which has been detected across a wide range of wavelengths from radio to TeV band \citep{2004RvMP...76.1143P, 2019Natur.575..455M}.

GRB prompt emission is created by energy dissipation as the relativistic jet accelerates particles via either internal shocks or magnetic reconnection \citep{2015AdAst2015E..22P}. These particles typically emit a non-thermal spectrum that is often dominated by synchrotron radiation \citep{2020NatAs...4..174B, 2020NatAs...4..210Z}. However, the detailed radiation physics of GRBs is not fully understood \citep{2015PhR...561....1K}. In practice, the prompt GRB spectrum is usually modelled phenomenologically as a ``Band'' spectrum \citep{Band:1993}. In addition, some spectra show additional features such as thermal components or multi-coloured blackbody peaks \citep{2017IJMPD..2630018P}, inverse Compton scattered components \citep{2001AdSpR..27..813D}, low energy spectral breaks \citep{2018A&A...616A.138O}, deviation from synchrotron spectra \citep{2011A&A...526A.110D}, etc. The physical/spectral parameters of prompt emission --- like the Lorentz Factor $\Gamma$, the peak energy \Ep, the isotropic equivalent energy $E_{\gamma,\mathrm{iso}}$, or the isotropic luminosity $L_{\gamma,\mathrm{iso}}$ --- show some correlations like the Amati correlation \citep{Amati:2006MNRAS}, which have been explored for understanding GRB properties as well as applying them for cosmology.

The interaction of the jet with the ambient medium gives rise to synchrotron emission, commonly known as the afterglow \citep{1997ApJ...476..232M, 1998ApJ...497L..17S, piran2005}. The afterglow is broadband and lasts much longer than the GRB: being visible for~\corr{hours to days in X-ray bands, days to weeks in optical}, and weeks to months at radio wavelengths. From the first afterglow detection by \bep~\citep[GRB~970228;][]{1997Natur.387..783C}, the understanding of afterglows has increased tremendously over the decades --- with a huge boost from the \neil~with its rapid response abilities~\citep{2004ApJ...611.1005G}. The afterglow emission is phenomenologically simple to model, and the flux $F$ is often fit by a simple power-law in both time and frequency, $F \propto t^{-\alpha} \nu^{-\beta}$. The temporal decay index $\alpha$ and spectral decay index $\beta$ typically follow the $\alpha - \beta$ closure relation predicted by the forward shock model~\citep{Zhang2021285294, piran2005}. Some GRB afterglows show features that provide insights into the physics of the source: for instance, jet breaks \citep{Rhoads_1999, 1999ApJ...519L..17S}, supernovae in long GRBs \citep{1998Natur.395..670G, article}, and flaring activity generated by various mechanisms~\citep{2005Sci...309.1833B, 2007ApJ...671.1921F}.

\thisgrb, first reported by \fermigbm~(GBM), is a long GRB with multiple pulses in the prompt emission (\citealt{2009ApJ...702..791M}). The optical afterglow was detected by the \ztf~ \citep[ZTF;][]{2014htu..conf...27B} and followed by multiple observatories in many wavebands. Here, we report our findings based on extensive follow-up of the source with multiple telescopes. The paper is organised as follows. In \S\ref{sec:obs}, we describe our observations and data reduction. We also list out public data from various sources that we have used in this work. \S\ref{sec:prompt} discusses the temporal and spectral characteristics of the prompt emission. In \S\ref{sec:afterglow} we undertake broadband modelling of the afterglow, showing clear evidence of late-time brightening. We conclude by discussing various causes for this in \S\ref{sec:discussion} and identifying the most plausible one.

\section{Observations and data analysis}\label{sec:obs}

In this section, we present the prompt and afterglow observations carried out by various space and ground-based telescopes.

\subsection{Prompt Emission}
\thisgrb was discovered by the \fermi~(GBM, \citealt{2009ApJ...702..791M}) at UT~2021-02-04~06:29:25 (hereafter, \fermiT). The source was first localised to RA = 109.1\degr, Dec = 9.7\degr\ (J2000) with a statistical uncertainty of 4.0\degr\  \citep{2021GCN.29390....1F}. The burst was also detected by \textit{Gravitational-wave high-energy Electromagnetic Counterpart All-sky Monitor} (GECAM-B, \citealt{2021GCN.29392....1L}), \kw~\citep{2021GCN.29415....1F}, and \asat~\citep{2021GCN.29410....1W}. The source localisation was refined by~\bal~ \citep{2021GCN.29391....1K}, and further by the Inter-Planetary Network (IPN) by using data from \fermi,~\integral,~\swift,~\kw,~and \mars-HEND~\citep{2021GCN.29408....1H}.

\begin{figure}
\centering
\includegraphics[width=\columnwidth]{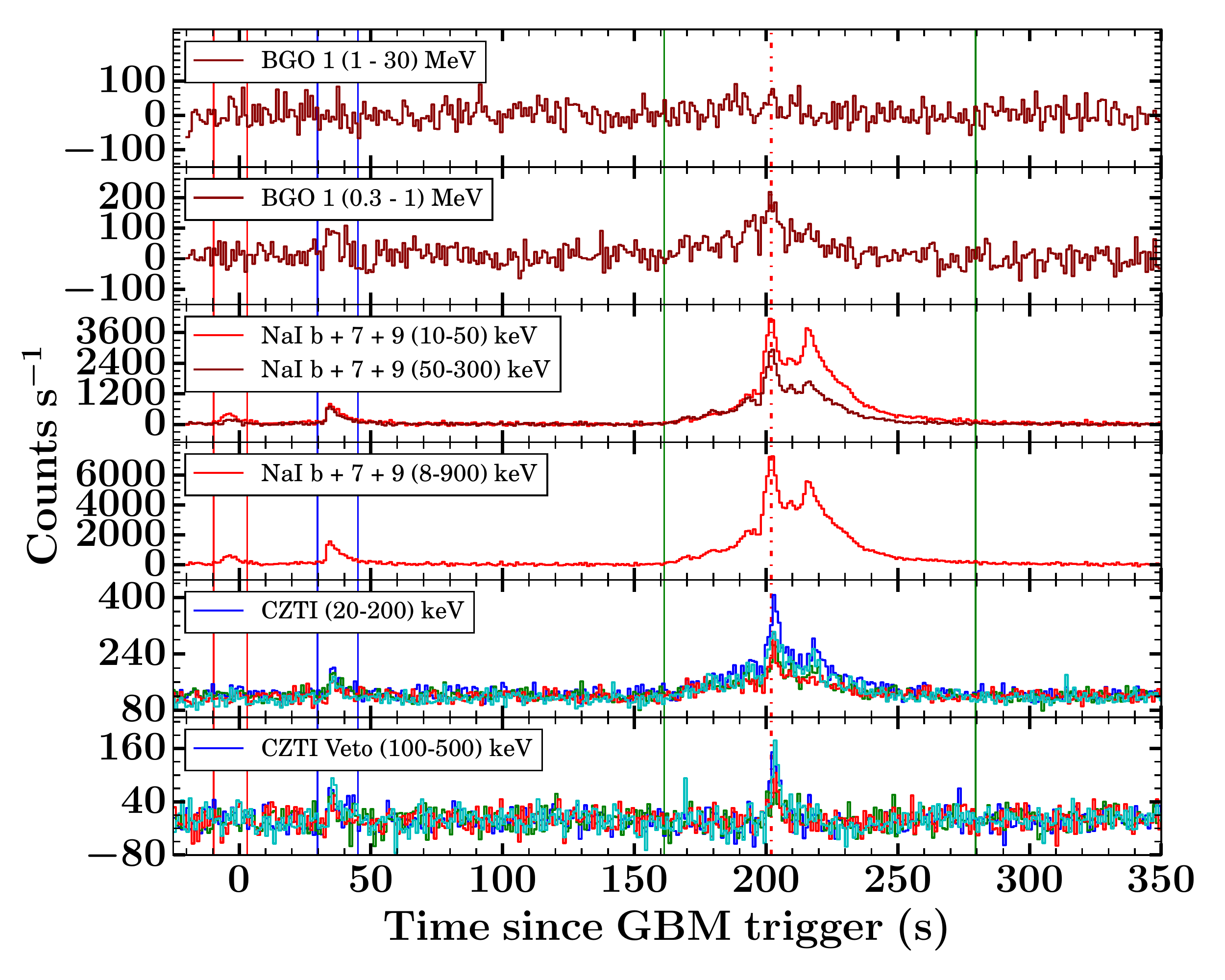}
\caption{\textit{Top four panels:} Energy-resolved \fermi-GBM prompt emission light curves (back-ground subtracted) of \thisgrb. The vertical red, blue and green lines indicate the duration of the first, second, and third episodes, respectively. The vertical dashed-dotted line indicates the peak used to calculate the isotropic luminosity of the burst.
\textit{Bottom two panels:} \asat-CZTI light curves in 20-200 \keV and 100-500 \keV energy range for \thisgrb. The four colours (blue, green, red, and cyan) correspond to data from four quadrants (A, B, C, and D) of the instrument. The GRB is detected more prominently in quadrants A and D due to the location of the GRB on the sky.} 
\label{fig:Prompt_LC}
\end{figure}

In this section, we focus on the analysis of data from \fermi~and \asat~(Figure~\ref{fig:Prompt_LC}).

\subsubsection{\bf \fermi-GBM}

We retrieved the \fermi-GBM data (the time-tagged event (TTE) mode) of \thisgrb from the \fermi~Science Support Center archives\footnote{\url{https://fermi.gsfc.nasa.gov/ssc/data/access/}}. We performed the temporal and spectral analysis of GBM data using three sodium iodide (NaI) detectors (NaI b, NaI 7, and NaI 9) and one bismuth germanate (BGO) detector (BGO 1). These detectors have following GRB observing angles NaI b: $\rm 25^\circ$ degree, NaI 7: $\rm 35^\circ$ degree, NaI 9: $\rm 47^\circ$, and BGO1: $\rm 25^\circ$, respectively. For the temporal analysis of \fermi-GBM data, we utilized \sw{RMFIT} version 4.3.2 software\footnote{\url{https://fermi.gsfc.nasa.gov/ssc/data/analysis/rmfit/}} and generated the prompt emission background subtracted light curve of \thisgrb in different energy ranges. Furthermore, we performed the spectral analysis of \fermi-GBM data using the Multi-Mission Maximum Likelihood framework \citep[\sw{3ML}\footnote{\url{https://threeml.readthedocs.io/en/latest/}}]{2015arXiv150708343V}. We performed time-integrated as well as time-resolved spectral analysis of GBM data to constrain the possible emission mechanisms of \thisgrb. We started the spectral modelling using traditional GRB model called \sw{Band} or \sw{GRB} function \citep{Band:1993}. In addition to \sw{Band} function, we explore various other possible models such as simple \sw{power-law} model, a power-law model with a high energy spectral cutoff (\sw{cutoffpl}), \sw{Black Body} function to search for photospheric signature in the spectrum, a power-law function with two sharp spectral breaks (\sw{bkn2pow}\footnote{\url{https://heasarc.gsfc.nasa.gov/xanadu/xspec/manual/node140.html}}), or combination of these models. We utilised the deviance information criterion \citep[DIC;][]{spiegelhalter2002bayesian} to find the best fit model. A more detailed methodology for GBM data analysis is discussed in \cite{2021MNRAS.505.4086G, 2022MNRAS.511.1694G}. 


\subsubsection{\bf \asat\ CZTI}

\asat\ Cadmium Zinc Telluride Imager \citep[CZTI;][]{2017JApA...38...31B} detected the second and third pulse of \thisgrb, with a total of 18141 photons: 94\% of which came from the brighter third pulse \citep{2021GCN.29410....1W,2020arXiv201107067S}. These two pulses were also clearly seen in the veto detectors. In detector coordinates, the GRB was incident from $\theta = 75.43\deg$ and $\phi =172.80\deg$: just 15\degr\ from the detector plane. CZTI can be used to measure the polarisation of GRBs by analysing two-pixel Compton events \citep{2015A&A...578A..73V,2014ExA....37..555C}. However, such measurements are robustly possible only for GRBs with $\theta < 60\deg$ \citep{2019ApJ...884..123C} --- ruling out the possibility of polarimetric studies of \thisgrb.

\subsection{Multiwavelength Afterglow}

The large 4\degr\ positional uncertainty in the \fermi\ localisation precluded prompt follow-up observations by most telescopes. However, \citet{2021GCN.29405....1K} used the wide-field ZTF and reported the discovery of a fast optical transient ZTF21aagwbjr/AT2021buv, a candidate afterglow for \thisgrb $\sim 38$~mins after the trigger. Subsequent follow-up observations by multiple telescopes verified the fading nature of this source and confirmed that it was indeed the afterglow of \thisgrb.

\subsubsection{X-ray afterglow}

Equipped with the precise afterglow position, the \neil~started Target-of-Opportunity observations of the \thisgrb field about $1.6\times 10^5$~s after the initial burst \citep{2021GCN.29412....1E}. The \swiftxrt~\citep[XRT;][]{2005SSRv..120..165B} detected an uncatalogued X-ray source at RA, Dec = $117.08071\deg, +11.40951\deg$ (J2000), consistent with the optical position. Multiple observations obtained till $3\times 10^5$~s after the burst confirmed the fading nature of this source.

We used the XRT online repositories by \citet{eva07, eva09} to retrieve the light curves\footnote{\url{https://www.swift.ac.uk/xrt_curves/}} and spectra\footnote{\url{https://www.swift.ac.uk/xrt_spectra/}}, respectively. We undertook spectral analysis with the X-Ray Spectral Fitting Package \citep[\sw{XSPEC};][]{1996ASPC..101...17A} version 12.10.1. The 0.3-10~keV spectra were modelled as a simple absorbed power-law (using the \sw{XSPEC} \texttt{phabs} model). For the time-averaged XRT spectrum (from \fermiT +1.61 $\times$ $10^{5}$ to \fermiT+ 1.73 $\times$ $10^{5}$ s), we get $\Gamma = 1.73^{+0.28}_{-0.26}$, and $N_\mathrm{H} = 6.43^{+5.40}_{-4.19} \times 10^{21}~\mathrm{cm}^2$.
In Table~\ref{tab:xrt}, we give the temporal evolution of XRT unabsorbed fluxes and photon indices (determined from the hardness ratio) obtained from the Swift Burst Analyser web page, supported by the UK Swift Science Data Centre.

\subsubsection{Optical afterglow}\label{subsec:optafterglow}
\citet{2021GCN.29405....1K} discovered the afterglow about 38~minutes after the initial burst. They also reported a non-detection of the same object in serendipitous observations of the field about 1.9~hours prior to their first detection (see \citet{2021ApJ...918...63A} and \citet{2022arXiv220112366H} for discovery details).
Follow-up observations obtained by various groups (see for instance Table~\ref{tab:public}) revealed that the source was indeed the fading afterglow of \thisgrb and measured the source redshift. We embarked on an extensive monitoring campaign using various telescopes~\corr{in the time interval between $\sim 0.03$ and $\sim 20$~days}, after the burst event.

We discuss our observations from four Indian facilities in this section and present a summary of data reported by other groups.

\begin{figure}
     \centering
     \begin{subfigure}[b]{0.4\textwidth}
         \centering
         \includegraphics[width=\linewidth]{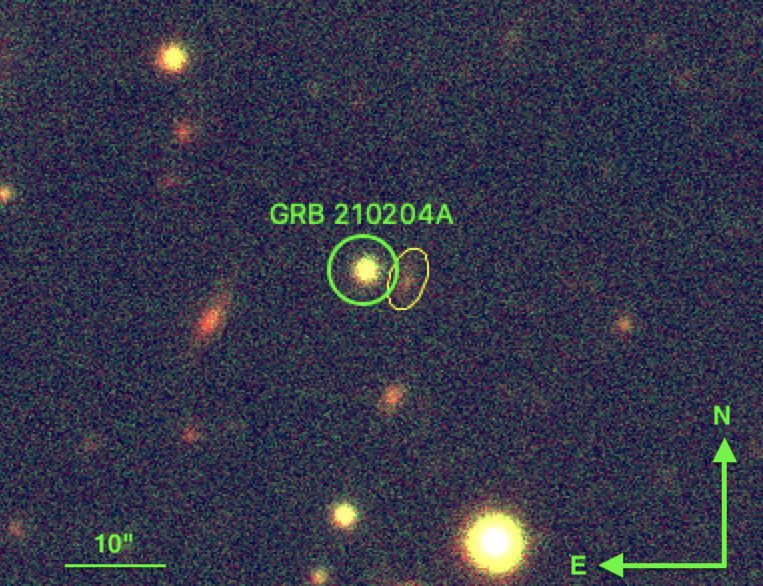}
         \caption{DOT image of GRB}
         \label{fig:DOT-grb}
     \end{subfigure}
     \hfill
     \begin{subfigure}[b]{0.4\textwidth}
         \centering
         \includegraphics[width=\linewidth]{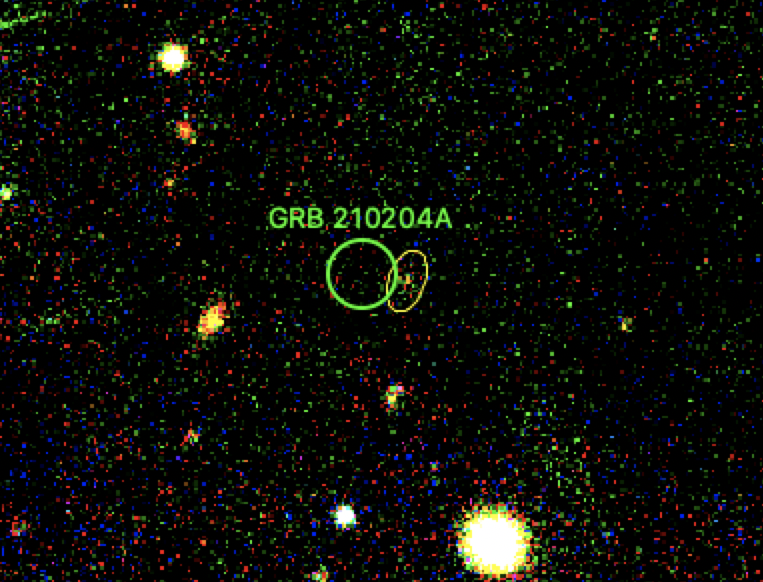}
         \caption{Pan-STARRS DR1 image}
         \label{fig:pan-grb}
     \end{subfigure}
        \caption{\thisgrb detection by the 4K $\times$ 4K CCD IMAGER mounted at the axial port of \doot~(DOT). (a) An image of GRB~210204A afterglow detection in DOT image. The position of afterglow is indicated by the green circle. There is a galaxy present at $\sim 4.4~\arcsec$ from the location of GRB (shown with a yellow ellipse), but it is unlikely to be the host (\S\ref{subsec:optafterglow}). (b) Snapshot of Pan-STARRS image of the same field is shown where no source is present at the position of afterglow.}
        \label{fig:optical_afterglow}
\end{figure}

\begin{figure*}
	\centering
	\includegraphics[width=\textwidth]{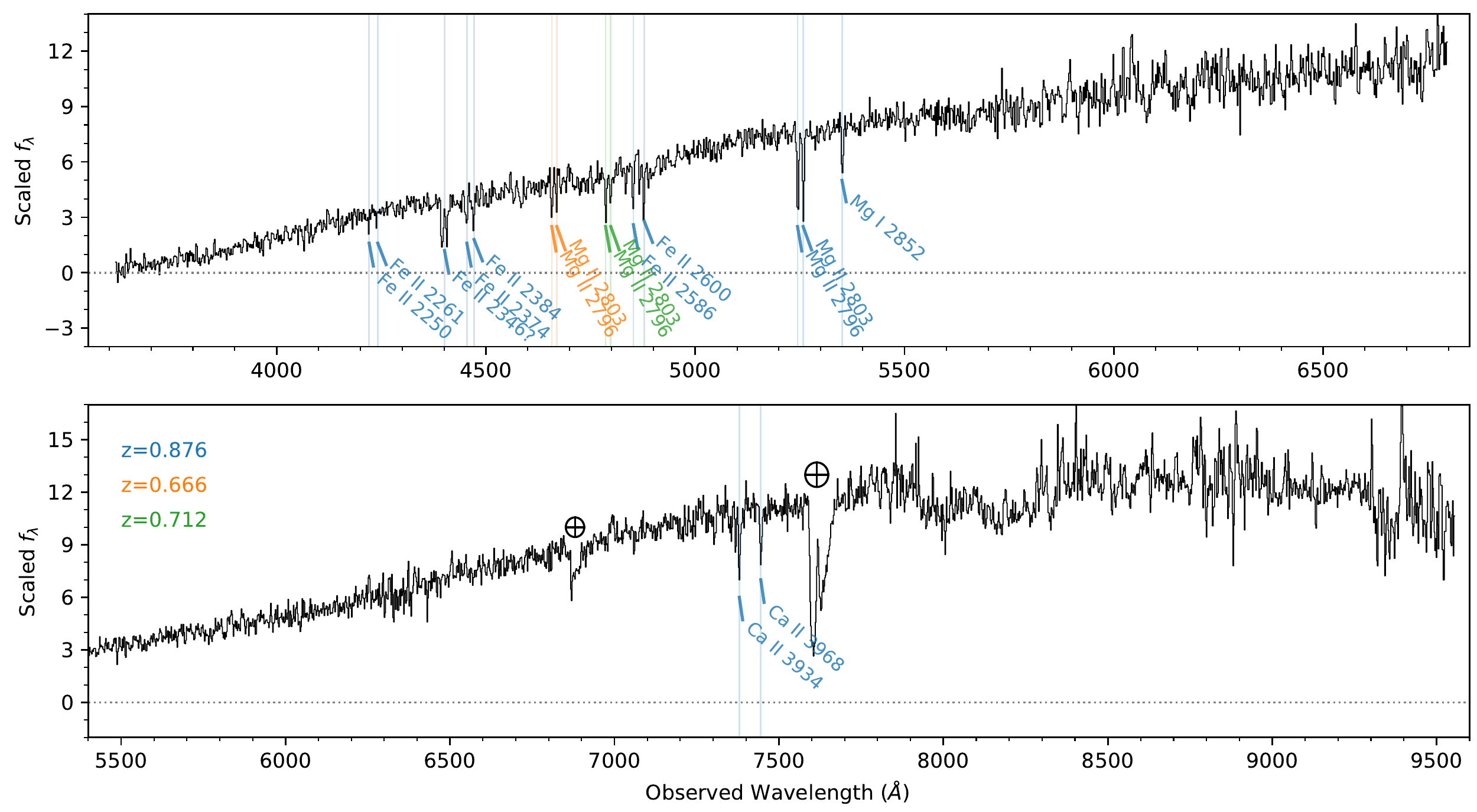}
	\caption{GMOS-S spectrum of ZTF21aagwbjr. The upper and bottom panel show the resulting spectrum in the blue and red gratings, respectively. We identify a number of strong, narrow absorption features of \ion{Fe}{II}, \ion{Mg}{II}, \ion{M}{I} and \ion{Ca}{II} at a common redshift of $z=0.876$ (blue notations). We identify two additional intervening absorbers, based on \ion{Mg}{II} $\lambda\lambda$2796, 2803 at $z=0.666$ (orange notations) and $z=0.712$ (green notations).
		\label{fig:spectrum}}
\end{figure*}

We followed up \thisgrb with the \git~(GIT), a 0.7~m telescope located at the Indian Astronomical Observatory, Ladakh. The telescope was equipped with a 2184$\times$1472 pixel {Apogee KAF3200EB} camera, giving a limited $11^\prime \times 7.5^\prime$ field of view. While poor observing conditions prevented immediate follow-up after the announcement of the ZTF discovery, our observations began on 2021 February 06, 2.36 days after the initial alert, and continued till March 1, 2021. Typical observations consisted of multiple 300~s exposures in the $\rm{r}^{\prime}$ filter, with each exposure having a limiting magnitude of $\sim20.5$~mag. We used the 2.0~m~\hct~(HCT) on three nights: 2021 February 07, 2021 February 12 and 2021 February 14, under proposal number HCT-2021-C1-P2. Data were obtained in Bessel V, R, and I filters (Table~\ref{table:phot-table-our}). The 3.6~m \doot~(DOT), located at the Devasthal Observatory of Aryabhatta Research Institute of Observational Sciences (ARIES), Nainital, India \citep{2019CSci..117..365S} was triggered under our ToO proposal number DOT-2021-C1-P62 (PI: Rahul Gupta) and DOT-2021C1-P19 (PI: Ankur Ghosh) for the follow up. We observed \thisgrb on multiple epochs with the 4K $\times$ 4K CCD IMAGER \citep{2018BSRSL..87...42P, 2021arXiv211113018K}. The first observations were obtained in BVRI filters \citep{2021GCN.29490....1G}, while data on subsequent nights were obtained in the SDSS~r filter. Further, we also obtained data with the 1.3~m \dfot~(DFOT) located at Devasthal observatory of ARIES, Nainital, India \citep{2011CSci..101.1020S} under our ToO Proposal ID DFOT-2021A-P6 (PI: Rahul Gupta). We obtained data in B, V, R, and I filters on 2021-06-06 (\fermiT + 2.4~d), and more data in R and I bands on 2021-02-13 UT.

Data obtained from all these facilities were reduced in similar manner using a python based reduction pipeline. Images were calibrated using bias and flat frames; pipeline made use of \sw{Astro-SCRAPPY}~\citep{2019ascl.soft07032M} python package to remove cosmic rays from the science images. Once the images were corrected for all artefacts, we solved the images for astrometry using astrometry.net \solf\ engine~\citep{Lang_2010} in offline mode. Sources in images were extracted in the form of a locally generated catalogue via \sxt~\citep{1996A&AS..117..393B}. \psfex~astromatic software~\citep{2011ASPC..442..435B} gave the PSF profile of the sources, which was used to get magnitudes of stars in the images. For images obtained in \textit{ugriz} filters, these magnitudes were cross-matched with Panoramic Survey Telescope and Rapid Response System (Pan-STARRS) DR1 catalogue~\citep{2016arXiv161205560C} and Sloan Digital Sky Survey (SDSS) DR12 catalogue~\citep{2015ApJS..219...12A} using VizieR to get the zero-points of the images. While, for \textit{BVRI} filter images, we used data from Sloan Digital Sky Survey (SDSS) \citep{2015ApJS..219...12A} and converted the magnitudes to VRI bands using Lupton (2005) transformations\footnote{\url{http://classic.sdss.org/dr4/algorithms/sdssUBVRITransform.html}} to estimate the zero-points. For later epochs where the afterglow was fainter, multiple exposures were stacked together using \swp~\citep{2002ASPC..281..228B}. Table~\ref{table:phot-table-our} lists the magnitudes with 1-sigma uncertainties. In case the source was not detected, we report 5-sigma upper limits.

In addition to the observations taken by our group, we also use publicly available data reported in Gamma-ray Coordination Network (GCN) by various groups. This set includes data from the ZTF published in \citep{2021ApJ...918...63A}, 1.6-m~\corr{\textit{AZT-33IK telescope}}\footnote{\url{http://en.iszf.irk.ru/Sayan_Solar_Observatory}}, 70~cm~\corr{\textit{AS-32 telescope}}~\citep{article_tels2},~\corr{\textit{Large Binocular Telescope}}~\cite{2010ApOpt..49D.115H}, 2.6-m \corr{\textit{Shajn Telescope}}~\citep{1976IzKry..55..208I} and the~\corr{\textit{AZT-20}}~at Assy-Turgen observatory\footnote{\url{https://fai.kz/observatories/assy-turgen}}. These data, along with the GCN references, are tabulated in Table~\ref{tab:public}.

 Figure \ref{fig:optical_afterglow} shows the detection of GRB afterglow with DOT (located by the green circle in the image). In DOT images, a galaxy is present $\sim 4.4~\arcsec$ away from the afterglow position. This transforms to a physical distance of $\sim 33$~kpc from GRB location, which is rather large for the galaxy to be the host of \thisgrb. Further, the photometric redshift of this galaxy is $z_{\mathrm{phot}} = 0.436$ makes it implausible to be the host of \thisgrb.

\corr{\textbf{Spectroscopy:}}\label{subsec:spec}
We triggered a long-slit spectrum of \thisgrb with GMOS-S under our ToO program GS-2021A-Q-124 (PI: A. Ho). The observation, conducted in the Nod-and-Shuffle mode with a 1\arcsec wide slit, started at 2021-02-06 01:19:09.2 UT, corresponding to 42.8 hours after the \fermi-GBM trigger.
We obtained $2 \times 450$\,s spectroscopic exposures with the B600 grating and $2 \times 450$\,s exposures with the R400 grating, providing coverage over the range 3620--9600\,\AA. Flux calibration was not performed. The spectrum was reduced using the \texttt{IRAF} package for GMOS. We identified a series of strong absorption features at $z=0.876$ superposed on a relatively flat, featureless continuum (Figure~\ref{fig:spectrum}). 

We also detected intervening absorption systems of \ion{Mg}{II} $\lambda\lambda$2796, 2803 at $z=0.666$ and $z=0.712$. This interpretation is consistent with that of \citet{2021GCN.29411....1I}, who obtained spectra using ESO VLT UT3 equipped with X-shooter spectrograph $~\sim 1.79$ days after the trigger. Their spectra spanned the wavelength range from 3000-21000~$\AA$ in which they report a few absorption lines of  Al II, Ca II, Fe II,  Mg I, Mg II, Zn II, Ca H and K detected at a common redshift of z = 0.876. They also detect three intervening Mg II absorbers at redshifts of z = 0.71, 0.66, and 0.57.

\subsubsection{Radio afterglow}\label{subsec:radio}

The \thisgrb event was triggered with the upgraded \gmrt~(uGMRT) at 2021 Feb 20.56 UT in band 5 (1000--1450~MHz). The observations were two hours in duration, including overheads using a bandwidth of 400 MHz. We use the Common Astronomy Software Applications \citep[CASA;][]{2007ASPC..376..127M} for data analysis.
The data were analysed in three major steps, i.e. flagging, calibration and imaging using the procedure laid out in \citet{2021ApJ...907...60M}. A source was clearly detected at the RA(J2000) = 07:48:19.34, Dec(J2000) = 11:24:33.91. This position is consistent with the position reported by ZTF for the GRB \citep{2021GCN.29405....1K}. Further follow-up observations were triggered on 2021 Mar 07.59 UT and 2021 Mar 09.56 UT in the uGMRT band 4 and band 5 respectively, 2 hours at each band including overheads. In both observation the source was detected with a resolution of $2.66\arcsec \times 1.74\arcsec$ and $6.87\arcsec\times2.05\arcsec$. Table \ref{tab:radio} lists the detailed radio followup information.

\section{Prompt emission}\label{sec:prompt}

We analyse the \fermi~data of the prompt emission to characterise \thisgrb and compare it with the overall GRB population.

\subsection{Spectral analysis of the complete GRB}

The prompt emission light curve of \thisgrb obtained using \fermi-GBM data shows three distinct episodes, separated by quiescent temporal gaps (see Figure~\ref{fig:Prompt_LC}). The first two episodes have relatively faint and simple fast rising and exponential decay profiles, but the third and brightest episode has rich sub-structure. The \tninty duration for the entire burst is 207.86 $\pm$ 0.06 s. The time-integrated~\corr{\textbf (the entire duration of the burst)} \fermi-GBM spectrum (from \fermiT-9.73 to \fermiT+279.55 s) could be best explained using traditional \sw{Band} plus \sw{Blackbody} model with following spectral parameters: peak energy (\Ep) = $146 \pm 14$~\keV, low energy spectral index $\alpha_{\rm pt} = -1.30\pm 0.07$, high energy spectral index $\beta_{\rm pt} = -2.39^{+0.17}_{-0.18}$ and temperature kT$_\mathrm{BB}$ $= 6.5 \pm 0.6$~\keV.

\begin{table}
\caption{Comparison between the characteristics of three episodes of \thisgrb. The quiescent duration between the first two episodes is $\sim26.2~$s, while that between the second and third episode is $\sim 116.2~$s. A blackbody component is needed only for the third episode (\S\ref{subsec:episodes}). All reported values are observer frame values. The total energy and luminosity were calculated using the source redshift $z = 0.876$. $\rm T_{90}$: Duration in 50--300~keV band; HR: ratio of the counts in 50 - 300 $\keV$ to the counts in 10 - 50 $\keV$; $\rm E_{p}$, $\alpha$, $\beta$: Band spectral fit parameters;
$\rm F$: Bolometric energy flux;
$\rm E_{\gamma, iso}$: Isotropic energy; $\rm L_{p, iso}$: Isotropic peak luminosity }
\label{tab:sub_bursts_properties}
\begin{center}
\begin{tabular}{|c|c|c|c|}
\hline
Characteristics & Episode 1 & Episode 2 & Episode 3 \\
\hline
\tninty (s) in 50 - 300 \keV & $12.04 \pm 0.02$& $12.81 \pm 0.04$ & $82.66 \pm 0.05$ \\
HR  & 0.41 & 0.79 & 0.57\\
\hline
\Ep (\keV) &$36 \pm 9$ &  $197 \pm 30$ & $146 \pm 9$  \\ 
$\alpha$ & $-0.96 \pm 0.36 $ & $-1.21 \pm 0.07$ & $-1.30 \pm 0.04$ \\
$\beta$ & $-2.14 \pm 0.14 $ & $-2.6 \pm 0.3$ & $-2.46 \pm 0.14 $\\
kT$_\mathrm{BB}$ & $\cdots$ & $\cdots$ & $6.39 \pm 0.40$ \\
\hline
$\rm F$ ($\rm 10^{-7} erg ~cm^{-2} ~s^{-1}$) & $1.6_{-1.13}^{+3.65}$& $3.8_{-0.7}^{+0.9}$ & $7.9 \pm 1 $\\
$\rm E_{\gamma, iso}$ ($\rm erg$) & $ 4.19 \times 10^{51}$ & $ 1.22 \times 10^{52}$ & $1.94 \times 10^{53}$ \\
$\rm L_{p, iso}$ ($\rm erg ~s^{-1}$) & $1.80 \times10^{51}$ & $4.44 \times10^{51}$ & $ 1.70 \times10^{52}$ \\
\hline
\end{tabular}
\end{center}\

\end{table}

\subsection{Episode-wise analysis}\label{subsec:episodes}

If we analyse the three pulses separately, we see that the \Ep values for the second and third pulses are higher (Table~\ref{tab:sub_bursts_properties}). We find that the band function gives acceptable spectral fits to the first and second episodes. The third episode is better fit by a power-law with two breaks (\texttt{bkn2pow}) or by a Band spectrum with an added blackbody component. The thermal component has a temperature of $6.4 \pm 0.4$ \keV. We use the Band + blackbody model in the rest of this section. We note that due to the lower intensity of the first two episodes, the data quality is not high enough to rule out such spectral features in them. 

\begin{figure}
\centering
\includegraphics[scale=0.36]{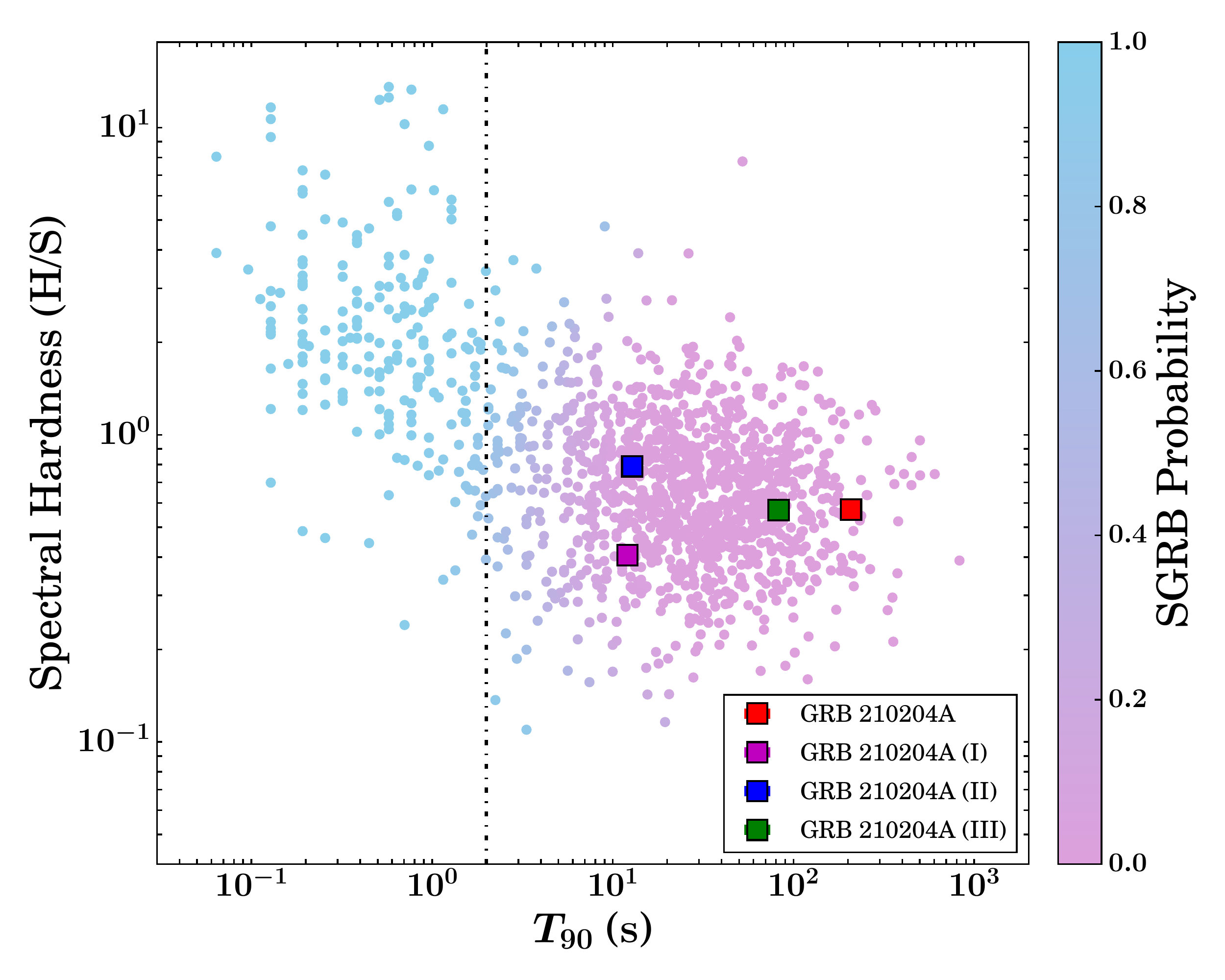}
\caption{Prompt emission \tninty-HR correlations: (a) Time-integrated (shown with a red square), and episode-wise (shown with magenta, blue, and green squares for the first, second, and third episodes, respective) \tninty-HR correlation for \thisgrb. We have also shown the data points for long and short GRBs taken from \protect\cite{2017ApJ...848L..14G}. The right side y scale shows the probability of short GRBs. The vertical black dashed-dotted line indicates the boundary between two classes of GRBs.}
\label{fig:T90HR}
\end{figure}

The presence of a thermal component along with a non-thermal component indicates a hybrid jet composition, including a matter-dominated hot fireball and a colder magnetic-dominated Poynting flux component for \thisgrb. The low energy spectral index values (Table~\ref{tab:sub_bursts_properties}) are within the range expected for synchrotron emission, $-3/2 < \alpha < -2/3$.

We calculated the \tninty values and the (50 -- 300~keV)/(10 -- 50~keV) hardness ratios for the entire GRB and the three episodes within it. Following \citet{Bhat:2016ApJS}, we estimated the errors in these by simulating 10,000 light curves by adding Poisson noise with mean equal to observed values and repeating these measurements on each simulated light curve. Figure~\ref{fig:T90HR} shows these values compared to the population of long and short GRBs --- we find that \thisgrb, as well as the three individual emission episodes within it, are all consistent with the ``long--soft'' GRB population.

\subsection{GRB global relationships}

The time-integrated rest-frame peak energy ($E_{\rm p,i}$) of the prompt emission spectrum of GRBs is correlated to the isotropic equivalent energy ($E_{\gamma,\rm iso}$), and this correlation is defined as Amati correlation \citep{Amati:2006MNRAS}. \cite{Basak:2013MNRAS} studied the episode-wise Amati correlation for a sample of \fermi-GBM detected GRBs with a measured redshift and confirmed that this correlation is more robust and valid for the episode-wise activity of GRBs. Recently, \cite{vchand} studied the Amati correlation for a sample of two-episodic GRBs and found that other than the first episode of GRB 190829A, each episode of two-episodic GRBs are consistent with the Amati correlation. In addition to GRB 190829A, a few other GRBs such as GRB 980425B, GRB 031203A, and GRB 171205 do not follow the Amati correlation. 

\begin{figure}
\centering
\includegraphics[scale=0.36]{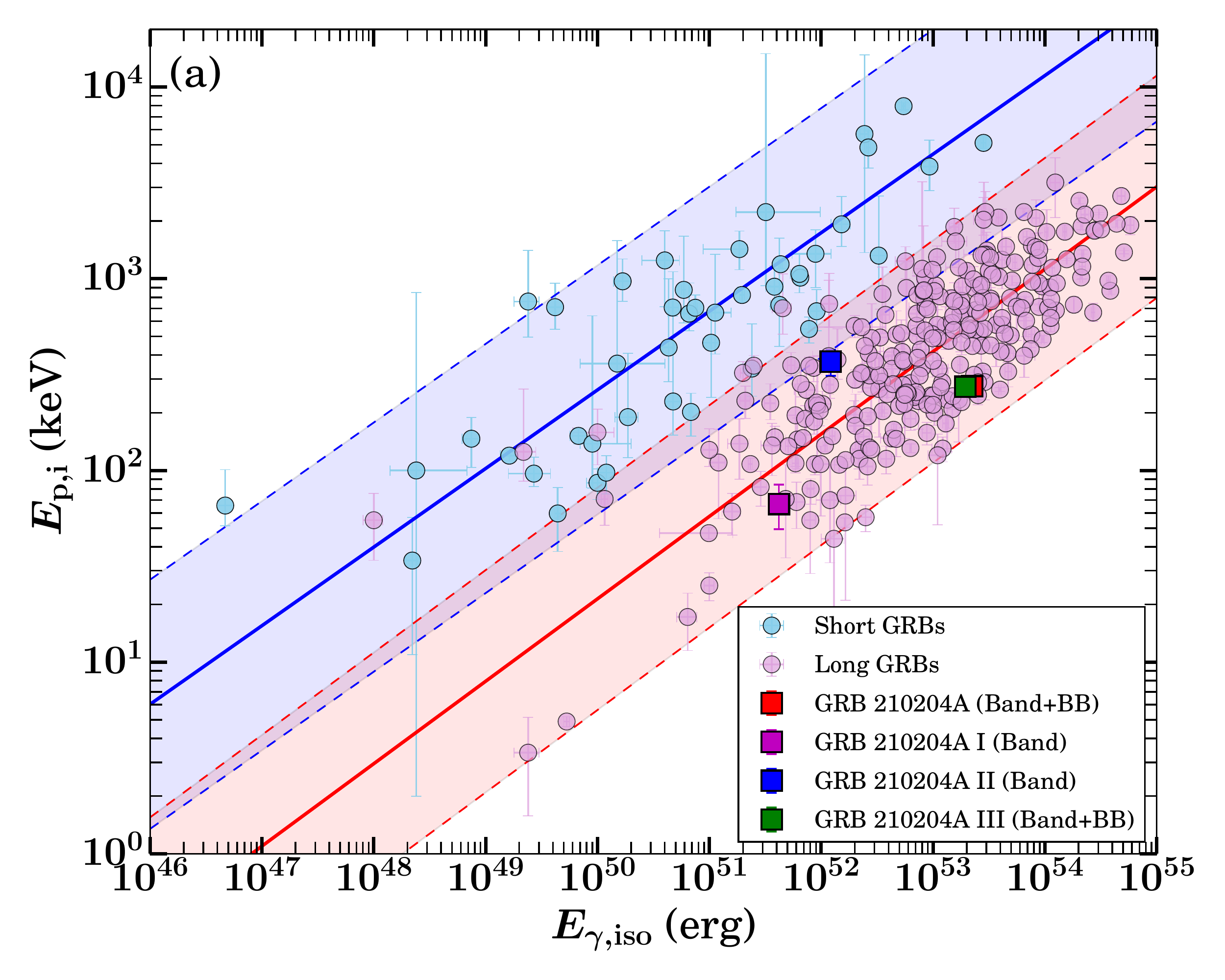}
\includegraphics[scale=0.36]{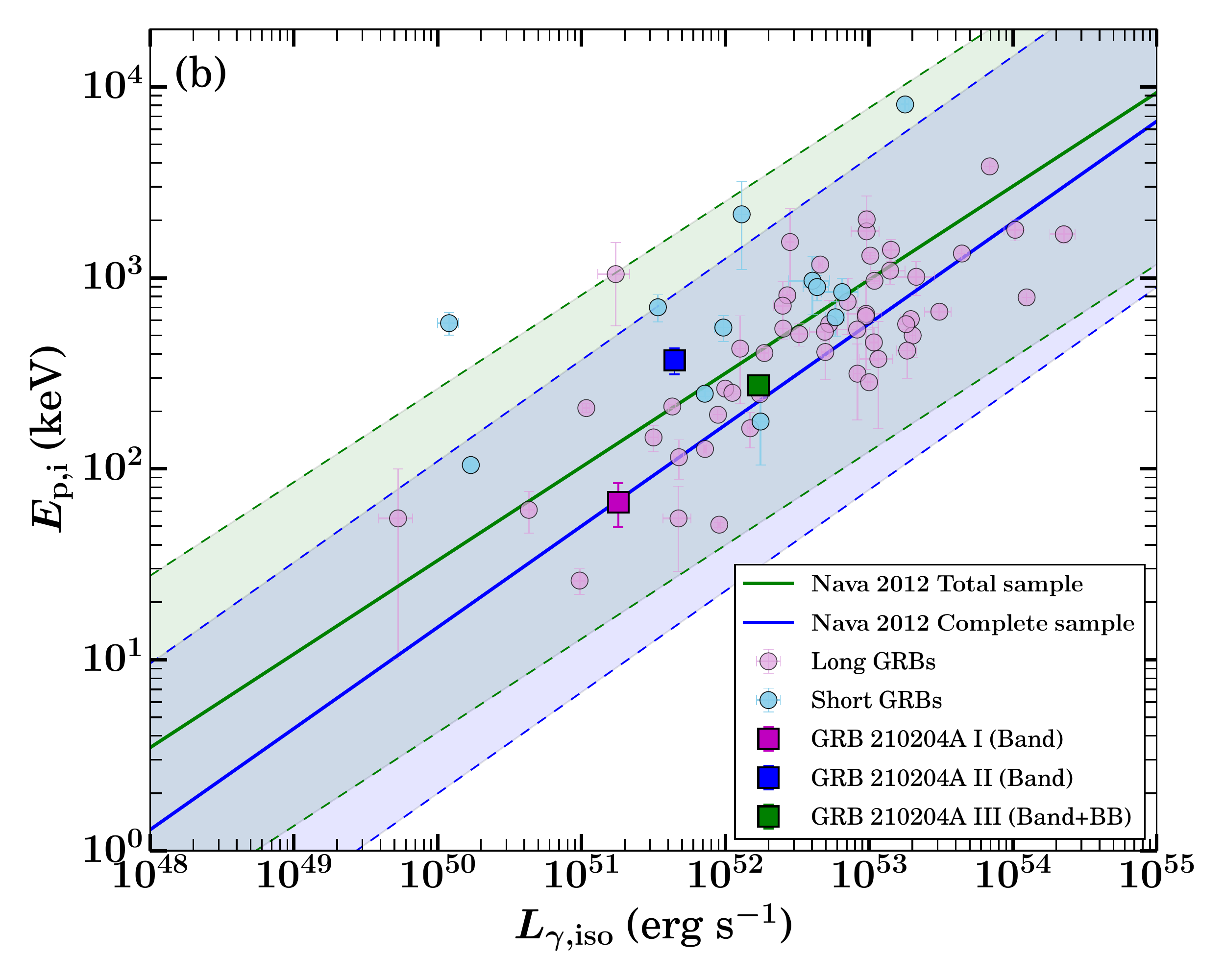}
\caption{Prompt emission Amati and Yonetoku correlations: (a) Time-integrated (shown with a red square), and episode-wise (shown with magenta, blue, and green squares for the first, second and third episodes, respective) Amati correlation for \thisgrb. Note that the \Ep values have been converted to the rest frame using the GRB redshift $z = 0.876$. We also show the data points for long and short GRBs taken from \protect\cite{2020MNRAS.492.1919M}. The red and blue solid lines show the best-fit line for long and short bursts, and the shaded regions represent the 2-$\sigma$ uncertainty region for both the populations of GRBs. (b) The episode-wise (shown with magenta, blue, and green squares for the first, second and third episodes, respective) Yonetoku correlation for \thisgrb. We also show the data points for long and short GRBs taken from \protect\cite{2012MNRAS.421.1256N}. The green and blue solid lines show the best-fit line, and the shaded regions represent the 3-$\sigma$ uncertainty region.}
\label{fig:Amati_Yonetoku}
\end{figure}

Another variant of Amati correlation is Yonetoku correlation which is the correlation between time-integrated rest-frame peak energy ($E_{\rm p,i}$) and isotropic peak luminosity $L_{\rm \gamma, iso}$ \citep{Yonetoku:2004ApJ}. These correlations have been utilised to classify individual episodes in GRBs with long quiescent phases. Figure~\ref{fig:Amati_Yonetoku} shows \thisgrb on the Amati and Yonetoku correlations. We find that the time-integrated, as well as individual episodes values, are consistent with the Amati correlation of typical long GRBs. Similarly, the $L_{\rm \gamma, iso}$, and $E_{\rm p,i}$ values for individual episodes are consistent with the Yonetoku correlation.

\section{Afterglow}\label{sec:afterglow}
The GRB blast-wave interacts with the circumburst medium giving rise to synchrotron emission, which is one of the primary signatures of standard GRB fireball model \citep{Granot_2002}. The electrons have a power-law energy distribution characterised by the index p, which results in the spectral energy distribution, which can be described as a series of multiple power-law segments. These segments join each other a particular frequencies known as break frequencies i.e. self absorption frequency  ($\nu_{\mathrm{sa}}$), cooling frequency ($\nu_{\mathrm{c}}$), and synchrotron frequency ($\nu_{\mathrm{m}}$). In optical and X-rays emission, synchrotron self-absorption does not play an important role and hence can be neglected. Depending on the ordering of two break frequencies $\nu_{\mathrm{c}}$ and $\nu_{\mathrm{m}}$, multiple spectral regimes are possible, which in turn govern the overall shape of the light curve as shown in \citet[Figure~1]{Granot_2002}. The temporal evolution of these frequencies along with the peak flux $F_\mathrm{\nu,max}$ determines the shape of the light curve. We first discuss the evolution of the afterglow, followed by calculation of these quantities after detailed analysis in section \ref{subsec:afterglowmcmc}.
 
\subsection{Afterglow evolution}\label{subsec:brokenpl}
 The optical light curve of \thisgrb shows typical afterglow behaviour --- a power-law decline that steepens at some point. The light curve is most densely sampled in the $r$ and $R$ bands; hence we use them for a first-cut analysis. Fitting a power-law to data from these bands from $\sim$1.4 to 8 days after the burst, we obtain indices $\alpha_{r} = 1.16 \pm 0.05$ and $\alpha_{R} = 1.17 \pm 0.04$. The fits are consistent with a constant offset in the two light curves, with $m_R = m_r - 0.24\pm0.03$. For the rest of the analysis, we scale the $R$ band data to the $r$ band by applying this offset to create a joint $r+R$ light curve.

\begin{figure}
\centering
\includegraphics[height=6cm,width=9cm]{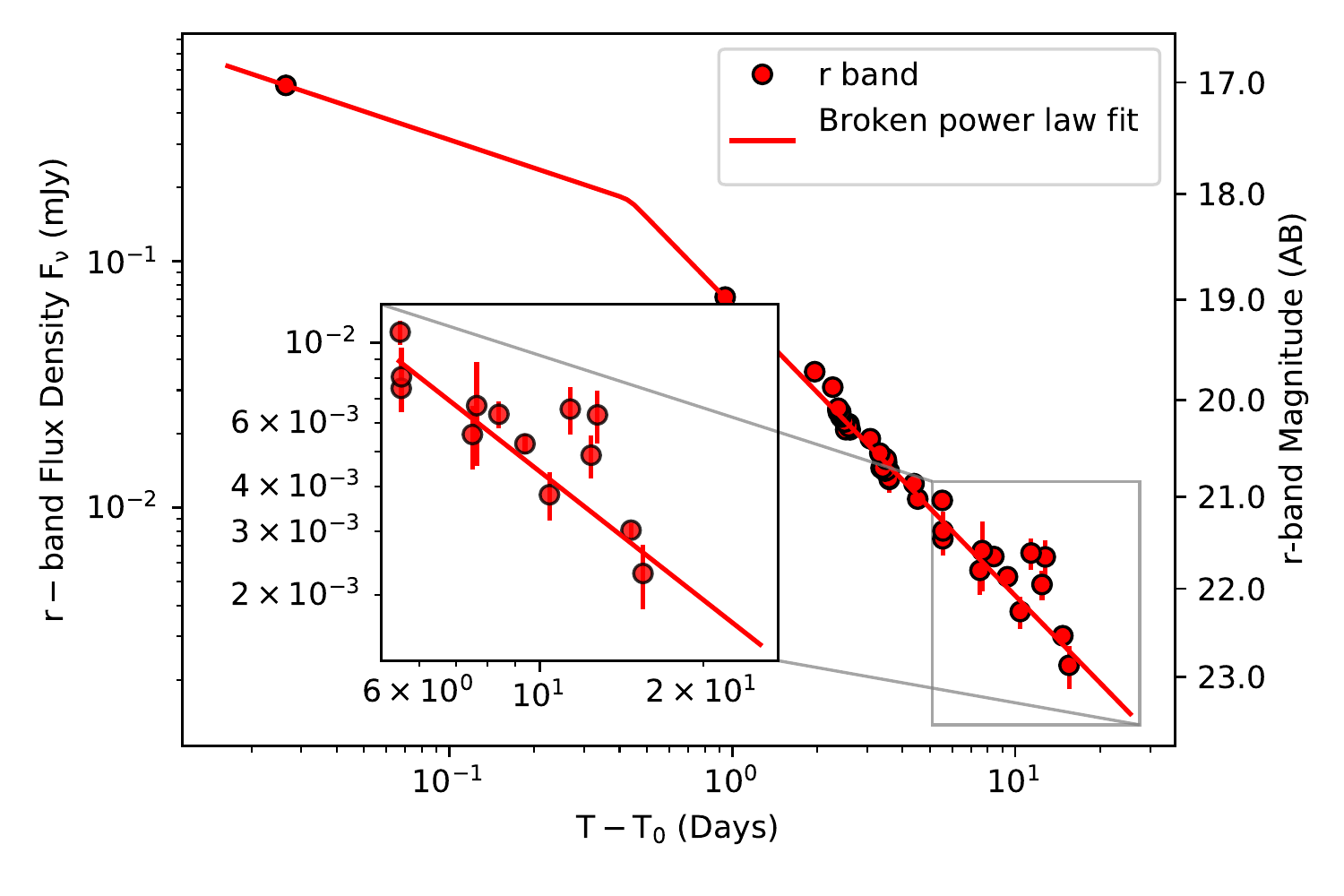}
\caption{Broken power law fit on r-band afterglow light curve. The red dots depicts the data points in r-band and the solid line is a broken power law fit on the data. The inset shows a zoomed in version of the light curve where it shows significant deviation from power-law around $\rm{T-T_{0}} \sim 10$~d.}  
\label{fig:rband-lc}
\end{figure}

The common light curve was used to fit a smoothly-joined broken power-law \citep{2015brokenpw} using the formula,
\begin{equation}
F_{\nu} = F_\mathrm{b} \left(\frac{(t/t_\mathrm{b})^{-y \alpha_1} + (t/t_\mathrm{b})^{-y\alpha_2}}{ 2}\right)^{-1/y}
\end{equation}
Here $F_{b}$ is flux at the break, $t_{b}$ is the time since the GRB at which the break-in power-law occurs, and the parameter $y$ ensures a smooth transition between the two power-law segments. The combined $r+R$ band light curve with the broken power-law fit is shown in Figure~\ref{fig:rband-lc}. We see a clear jet break early in the light curve, with a shallow temporal power-law index $\alpha_1\sim$ 0.33 at initial times. Due to the free smoothness parameter ($y$), and limited $r$ and $R$ data in early days ($t < 1$~day), the break time is rather poorly constrained to be $t_b = 0.37 \pm 0.30$~d (1-$\sigma$ error). After the jet break within the first day, the decline is steeper with power-law index $\alpha_2 = 1.18 \pm 0.03$.

The light curve shows a significant deviation from the power-law fit at $\rm{T-T_{0}} \sim 10$~days, seen clearly in the inset in Figure~\ref{fig:rband-lc}. In order to understand these deviations, we first undertake a detailed broadband fit while excluding these days from the data in \S\ref{subsec:afterglowmcmc}, and revisit the residuals in \S\ref{subsec:flarefit}

\subsection{Broadband afterglow modelling}\label{subsec:afterglowmcmc}

\begin{figure*}
    \centering
    \includegraphics[width=0.9\textwidth]{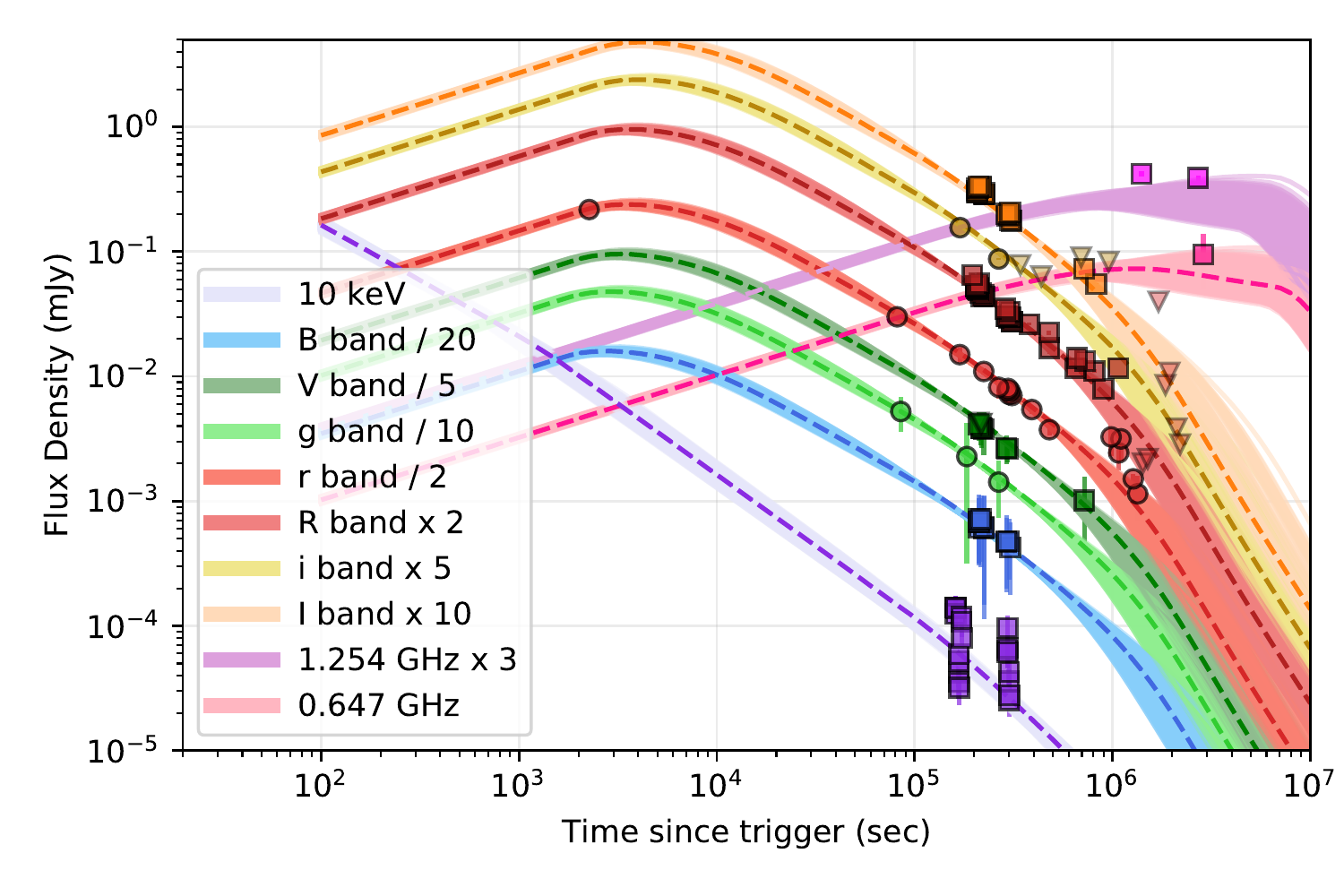}
    \caption{A multi-band light curve of \thisgrb afterglow. The multi-band light curve of the \thisgrb afterglow was fitted to the data using the $\af$ package integrated with the $\af$ with \sw{EMCEE} python package for Markov Chain Monte Carlo (MCMC). Dotted lines show the best fit light curve to each band, and the light coloured band around the dotted line indicates the  16\% and 84\% quantiles uncertainty in the fitting. }
    \label{fig:grb-mcmc}
\end{figure*}

 We performed a detailed analysis of the multi-wavelength light curve of the \thisgrb afterglow using the \af\ package \citep{afterglowpy-ryan, 2021NatAs...5..917A}. The \af\ python package is an open-source computational tool to compute the afterglow light curves for the structured jet. It has the capabilities to provide light curves for arbitrary viewing angles. We integrated the \af\ with \sw{EMCEE}~\citep{emcee} python package for Markov Chain Monte Carlo (MCMC) routine \citep{1953JChPh..21.1087M} to generate the posterior of parameters thanks to the fast light curve generation of $\af$. We included all radio and X-ray data in our modelling but limited our optical data to $T - T_0 < 8$~d, in order to avoid the ``brightening'' seen in \S\ref{subsec:brokenpl}.

\begin{table}
\caption{Posterior sampling using MCMC and \af}
\begin{center}
\begin{tabular}{c c c c c}
\hline
Parameter & Unit & Prior Type & Posterior  & Parameter Bound \\
\hline
${\theta_{\mathrm{obs}}}$ & rad & $\sin{({\theta}_{\mathrm{obs}})}$  & $0.010^{+0.002}_{-0.002}$  & [0.001, 0.8]  \\
$\rm{log}_{10}(E_0)$ & erg &  uniform& $54.06^{+0.03}_{-0.03}$ & [48, 56]\\
$\theta_{\mathrm{core}}$ & rad & uniform & $0.024^{+0.003}_{-0.002}$ & [0.01, $\pi/2$] \\ 
$\log_{10}(n_{\mathrm{0}})$ & cm$^{-3}$ &  uniform  & $-5.67^{+0.13}_{-0.16}$  & [-6, 100] \\
$p$ & - &  uniform & $2.18^{+0.026}_{-0.026}$ & [2.0001, 4] \\
$\epsilon_{\mathrm{e}}$ & - & - & 0.1 & - \\ 
$\log_{10}(\epsilon_{\mathrm{B}})$ & - &  uniform & $-0.86^{+0.17}_{-0.15}$  & [-6, 0] \\ 
$\xi$ & -& -& 1 & - \\
\hline

\end{tabular}
\end{center}
\label{tab:mcmc_post}
\end{table}

We used the TopHat jet model in $\af$ which performs artificial light curve modelling using a standard synchrotron fireball model. The temporal decay index $\alpha$ can be used to calculate the electron power-law index $p$ for the circum-burst medium using the closure relation $\alpha = 3(p-1)/4$~\citep[Table 2]{2020ApJ...900..176L}. For constant density Inter-Stellar Medium (ISM), the optical and X-ray decays yield $p_\mathrm{ISM,o} \sim 2.56 $ and $p_\mathrm{ISM,x} \sim 2.47$. On the other hand, for a wind-like medium, $\alpha = (3p-1)/4$, corresponding to unusually low values $p_\mathrm{wind,o} \sim 1.91$ and $p_\mathrm{wind,x} \sim 1.80$. We can also calculate that spectral index is $\beta \sim 0.75$ using the optical and X-ray fluxes, which in turn gives $p \sim 2.5$: consistent with the constant density ISM case. Hence, we proceed with detailed analysis assuming a constant density ISM.

Assumption of constant density medium (synchrotron Self Absorption is not important as discussed in Section \ref{sec:afterglow}) may cause disagreements between the model and the radio data. However, we find that the results do not significantly change whether we include radio in the fits. The MCMC routine was run to fit for angle between the jet axis and the observer ($\theta_\mathrm{obs}$), the total energy of the jet ($\rm{log_{10}}(E_{0})$), the half opening-angle of the jet ($\rm{\theta}_{core}$), the circumburst density ($\rm{log_{10}}(n_{0})$), the power-law index for the electron energy distribution ($p$), and the fraction of energy in electrons and the magnetic field ($\rm{log_{10}}(\epsilon_{e})$ and $\rm{log_{10}}(\epsilon_{B})$ respectively). The priors and bounds used for each parameters are shown in Table \ref{tab:mcmc_post}. We assumed a uniform distribution for $\theta_\mathrm{core}$, but we took the prior for the observer angle to be uniform in $\sin \theta_\mathrm{obs}$ to account for the uniform random orientations of sources in space \citep[see for instance][]{troja2018}. The exponent $p$ was assumed to be distributed uniformly in the semi-open interval $(2, 4]$: implemented practically as a uniform distribution in $[2.0001, 4]$. Finally, log-uniform priors were used for $\rm{E}_{0}$, $\rm{n}_{0}$, and $\epsilon_{B}$. Our preliminary fits showed that the data could not constrain $\xi$ and $\epsilon_{e}$ well, so we fixed them at nominal values of 1 \citep{afterglowpy-ryan} and 0.1 \citep{2002ApJ...571..779P, 2021arXiv211111795G} respectively. The source redshift was held fixed at 0.876 as discussed in \S\ref{subsec:spec}. Inputs for the fitting were the time since an event, observation frequency, measured flux, and the flux uncertainty. \af\ was used to generate models for various values of the input parameters, which were then compared to the observed data. The best-fit parameters and the confidence intervals were evaluated by maximising the likelihood of the model fits to the observations.

The one and two dimensional marginal posterior distribution resulted from the routine are shown in Figure~\ref{fig:corner}. For each parameter distribution median posterior and 16\% and 84\% quantiles are plotted at the top of panel, which we also quote as the parameter bounds here. The model constrained the jet isotropic energy to be $10^{54.06 \pm 0.03}$~ergs, consistent with typical long GRB afterglows \citep{2012MNRAS.423.2627W}. The jet structure parameter $(\rm{\theta}_{core})$ and viewing angle $(\rm{\theta}_{obs})$ were constrained at $0.024^{+0.003}_{-0.002}$~rad, $0.010\pm0.002$~rad respectively. From the values of $\rm{\theta}_{obs}$ and $\rm{\theta}_{core}$ it is evident that the jet is seen on axis (${\theta}_\mathrm{obs} < {\theta}_\mathrm{core}$).
 
The best fit model generated from afterglow + MCMC routine fit is shown in Figure~\ref{fig:grb-mcmc}. Markers denote observed flux densities, and in several cases, the error bars are smaller than the marker size. Dashed lines show the light curves in various bands, generated using median values from parameter distribution from MCMC routine. The shaded coloured bands show 16\% -- 84\% uncertainty regions around the median values. The fit indicates that the optical light curve would have risen at very early times, which is plausible based on the values of the synchrotron break frequencies at that time --- however, we do not have any observational data to constrain this. Note that the figure shows all data, even the points at $T - T_0 > 8$~d that were excluded from the fit. We can clearly see that the re-brightening episodes have statistically significant deviations from the fit values and are indeed astrophysical in nature.

Next, we tried to estimate the break frequencies and peak time of light curve. For this we consider a spherical shock propagating in a constant density ($n$) medium. The hydrodynamic evolution of this shock can either be radiative or adiabatic, which affects the late-time light curve behaviour. Following \citet{1998ApJ...497L..17S}, if we model the flux in decaying part of light curve as $F \sim t^{-\beta}$, then the decay index can take two values in the adiabatic case: $\beta_1 = 3(p-1)/4$ or $\beta_2 = 3p/4 -1/2$. Using value of $p$ from Table \ref{tab:mcmc_post}, we get $\beta_{\mathrm{1}} = 0.88$, $\beta_{\mathrm{2}} = 1.13$. On the other hand, $\beta \sim 3/7$ for fully radiative evolution.
In \S\ref{subsec:brokenpl}, we measured this late time decay index to be $\alpha_2 = 1.18 \pm 0.03$: close to the radiative $\beta_2$ calculated here. We conclude that the hydrodynamic shock evolution is adiabatic in nature. Hence, the equations governing shock parameters in the observers' frame are \citep{1998ApJ...497L..17S}:

\begin{eqnarray}
\nu_{\mathrm{c}} &=& 2.7 \times 10^{12} * (1+z)^{-1/2} \epsilon_{\mathrm{B}}^{-3/2} E_{52}^{-1/2} n_{0}^{-1} t_{\mathrm{d}}^{-1/2}~\mathrm{Hz} \label{eq:nuc}\\
\nu_{\mathrm{m}} &=& 5.1 \times 10^{15} *  (1+z)^{1/2} \left( \frac{p-2}{p-1} \right) ^{2} \epsilon_{\mathrm{e}}^{2} \epsilon_{\mathrm{B}}^{1/2} E_{52}^{1/2} t_{\mathrm{d}}^{-3/2}~\mathrm{Hz} \label{eq:num}\\
t_{\mathrm{m}} &=& 2.98 * (1+z)^{1/3} \left( \frac{p-2}{p-1} \right )^{4/3} \epsilon_{\mathrm{e}}^{4/3} \epsilon_{\mathrm{B}}^{1/3} E_{52}^{1/3} \nu_{\mathrm{15}}^{-2/3}~\mathrm{days} \label{eq:tm}\\
t_{\mathrm{0}} &=&1.89  \times 10^{3} * (1+z) \left( \frac{p-2}{p-1} \right )^{2} \epsilon_{\mathrm{e}}^{2} \epsilon_{\mathrm{B}}^{2} E_{52} n_{0}~\mathrm{days} \label{eq:t0}
\end{eqnarray}

\begin{figure}
    \centering
    \includegraphics[width=0.9\columnwidth]{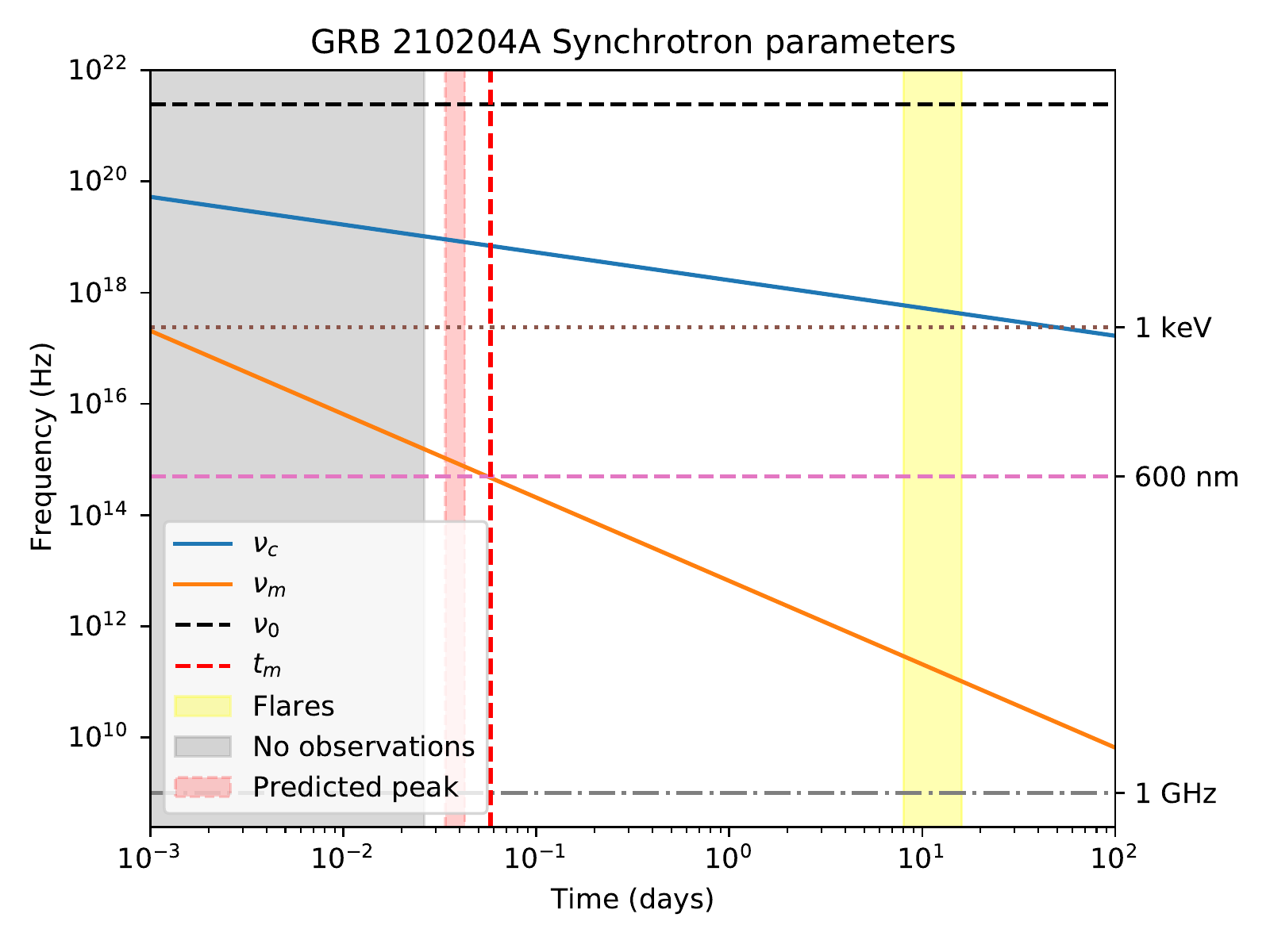}
    \caption{The interplay of break frequencies with time. The blue and orange solid line represents variation of $\nu_{c}$ and $\nu_{m}$ respectively with time. $\nu_{m}$ passes through the optical frequency (horizontal magenta dashed line) at $t = t_{m}$ (vertical dashed red line). Grey shaded region shows epochs where no observations were made. The yellow shaded region depicts flaring event. The narrow light red region at $\sim 3 \times 10^3$~s is where the light curve is predicted to be peaked through afterglowpy modelling of light curve.}
    \label{fig:break-freq}
\end{figure}

Here, $t_{d}$ is time in days since the trigger, $n_{0}$ is the Interstellar Medium (ISM) density in units of $\mathrm{cm}^{-3}$, $E_{52} = E_{0}/10^{52}$~ergs, $\nu_{\mathrm{15}} = \nu/10^{15}~\mathrm{Hz}$ and $t_{\mathrm{m}}$ is peak time. At  $t = t_0$, equation 2 and 3 satisfy $\nu_{0} = \nu_{c}(t_{0}) = \nu_{m}(t_{0})$~\citep{1998ApJ...497L..17S} where $\nu_{0}$ is called critical frequency. $t = t_0$ is the time at which the ejecta transitions from fast cooling to slow cooling phase. Using best--fit values from Table \ref{tab:mcmc_post}, Equation~\ref{eq:t0} yields $t_0 \sim 3.85 \times 10^{-6}$~days --- showing that \thisgrb\ transitioned to the slow cooling phase at very early times. This in turn gives a ``critical frequency'' $\nu_{\mathrm{0}} = 2.21 \times 10^{21}$~Hz which lies in $\gamma-\mathrm{ray}$ frequency range as shown by horizontal black dashed line in Figure \ref{fig:break-freq}, suggesting that the light curve shown in Figure \ref{fig:grb-mcmc} is a low-frequency light curve \citep{1998ApJ...497L..17S}. The optical light curve will peak when the synchrotron frequency ($\nu_{\mathrm{m}}$) passes through optical bands at $t_{\mathrm{m}} = 0.055~\mathrm{days} = 4.75 \times 10^{3}~\mathrm{s}$, in agreement with the \af\ fits for \thisgrb shown in Figure \ref{fig:grb-mcmc}. However, we lack sufficient early-time data to constrain such a rise. On the other hand, the cooling frequency at $\sim 1$~day $\nu_{\mathrm{c,t=1}} = 1.68 \times 10^{18}$~Hz, lies in X-ray bands ($E \sim 7$~keV) which accounts for different decay in X-ray and optical bands at early times.

\subsection{Quantifying the re-brightening}\label{subsec:flarefit}

\begin{figure*}
    \centering
    \begin{subfigure}{0.32\linewidth}
    \centering
    	\includegraphics[width=\columnwidth]{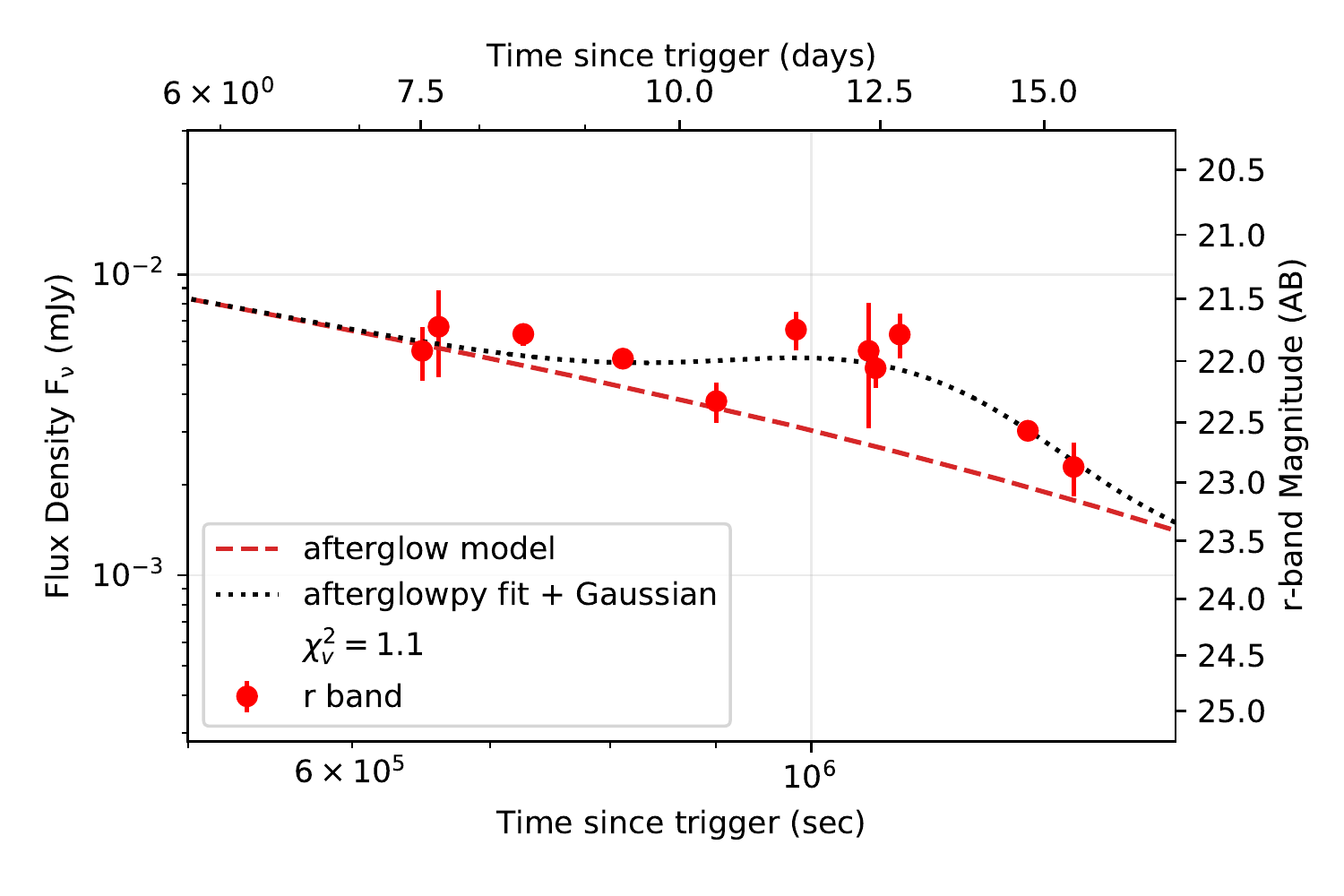}
    	\caption{1D Gaussian fit}\label{fig:gauss}
    \end{subfigure} \hfill
    \begin{subfigure}{0.32\linewidth}
    \centering
    	\includegraphics[width=\columnwidth]{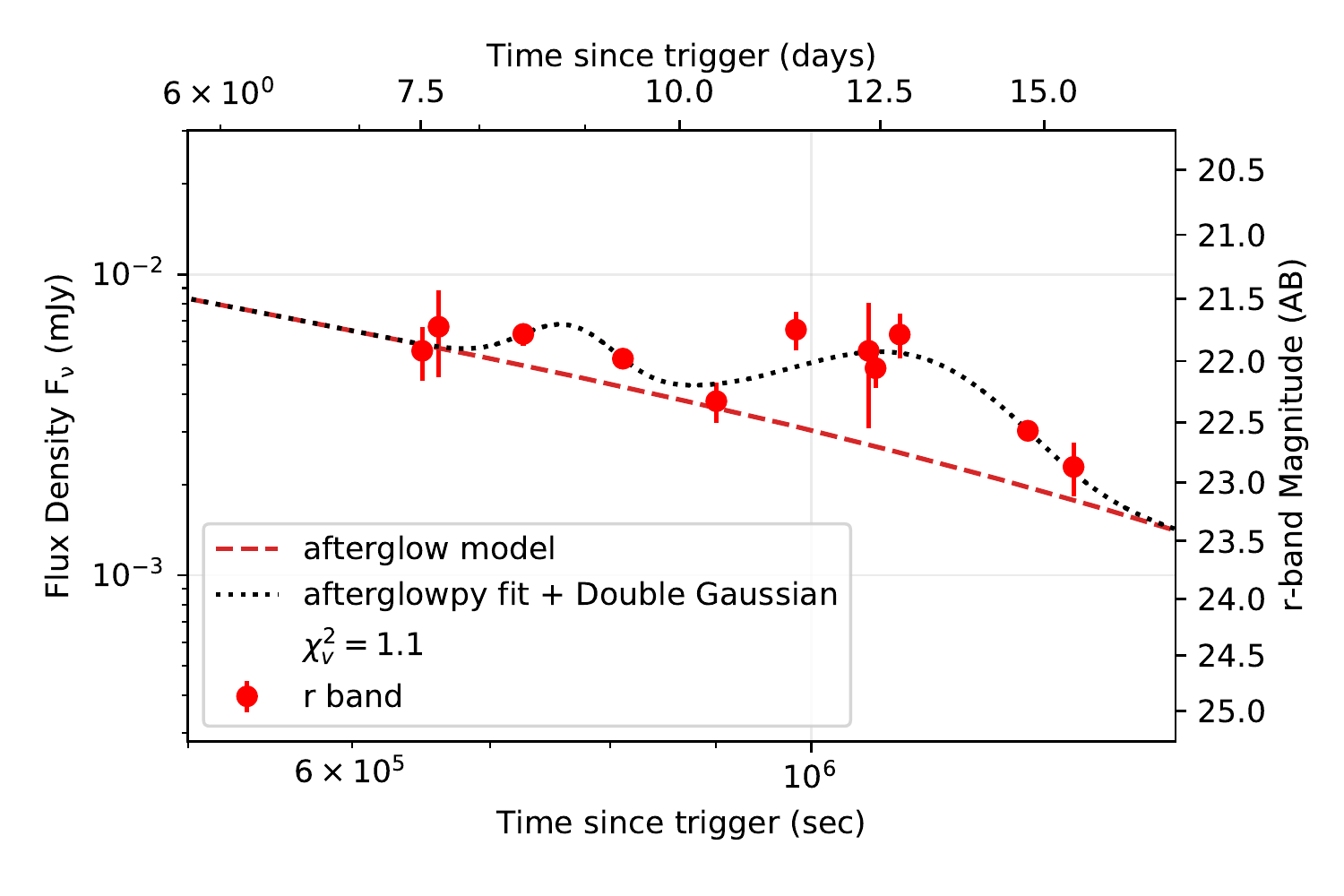}
    	\caption{1D Double Gaussian fit}\label{fig:double-gauss}
    \end{subfigure} \hfill
    \begin{subfigure}{0.32\linewidth}
    \centering
    	\includegraphics[width=\columnwidth]{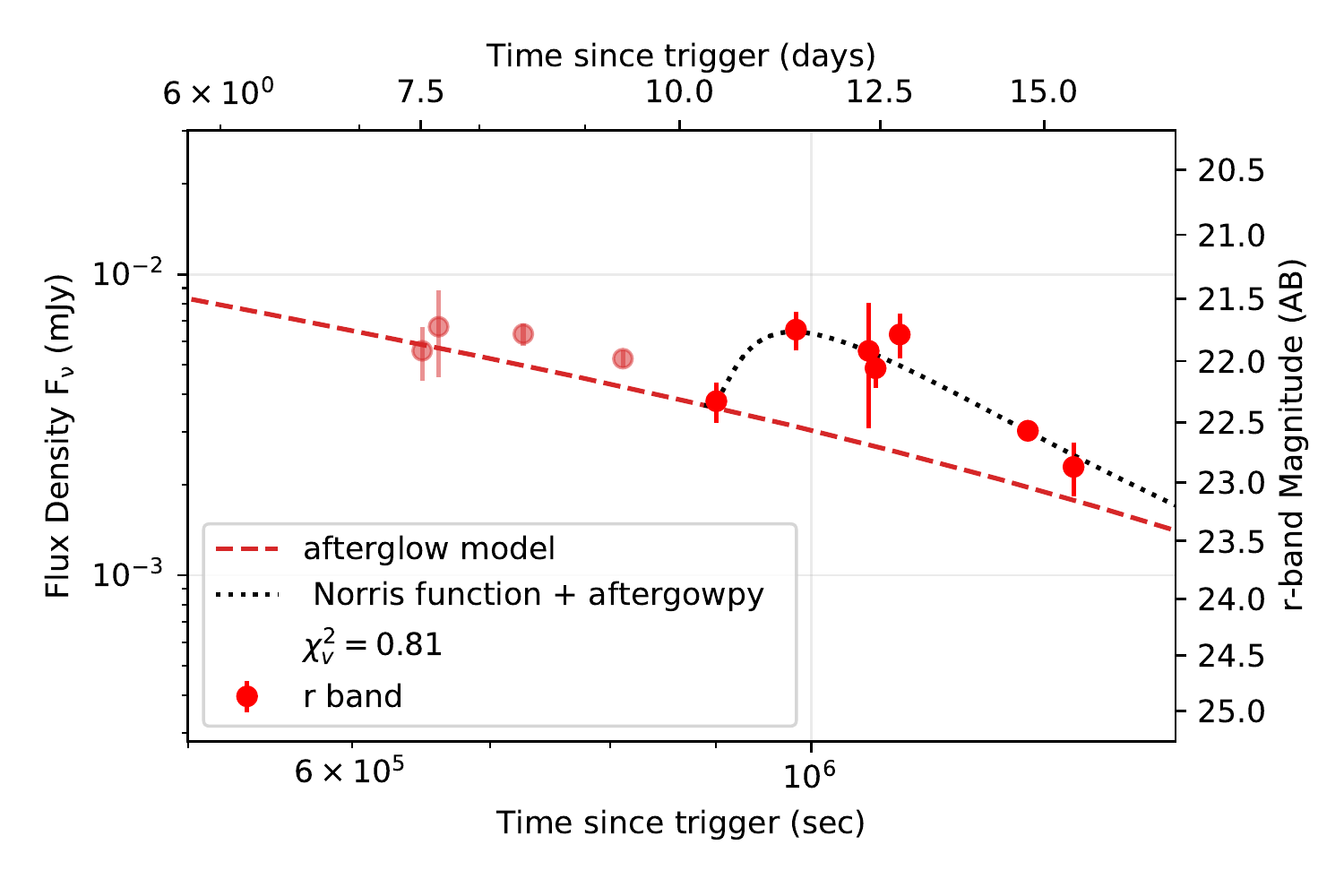}
    	\caption{Norris function fit} \label{fig:norris_fit}
    \end{subfigure}
     \caption{Late-time excessive emission fitting with various models. a) A simple Gaussian fitting to the peak. b) A double Gaussian fit to the late-time variability. c) Norris function fitted on the two peaks for flaring. The solid dark markers shows the data used for fitting for a particular function.}
     \label{fig:bumpt-fit}
\end{figure*}

Armed with our best-fit model for the afterglow, we revisit the re-brightening episode discussed in \S\ref{subsec:brokenpl}. Figure~\ref{fig:rband-lc} shows that these episodes occur only for a few nights, after which the data seem to return to the original power-law decay. We verified this starting with detailed quality checks on our data in this time range: including visual inspection, checking the stability of light curves of nearby stars, and re-checking the zero points. We find that the photometric measurements are robust, and the data indeed are brighter than the level expected from the afterglow model.

Next, we fit simple models to these episodes to measure their properties. In all fits, we use the nominal afterglow light curve from \S\ref{subsec:afterglowmcmc} as a ``background'' (red dashed lines in Figure~\ref{fig:bumpt-fit}), and add various ``flare'' models to this. We start with a simple Gaussian in flux density space: $F = F_0 \exp [ (t - t_\mathrm{peak})^2 / (2 \sigma^2) ]$, where $F_0$ is the peak flux density, $t_\mathrm{peak}$ is the time of the peak, measured from the GRB $T_0$, and $\sigma$ is the duration parameter. We obtain best-fit values as $F_0 = 0.0024 \pm 0.0006$~mJy, $t_\mathrm{peak} = (105 \pm 3) \times 10^4~\mathrm{s} = 12.1 \pm0.3$~d and $\sigma = (17.2 \pm 3.3) \times 10^4~\mathrm{s} = 2.0 \pm0.4$~d. This corresponds to an overall fluence of $7.87 \times 10^{-10}$~\fluence. However, the quality of the fit is not very good (Figure~\ref{fig:bumpt-fit}a), as the photometry is close to the predicted light curve till $\sim10$ days, and rises strongly after that. Hence, we fit two gaussians to the data, as shown in Figure~\ref{fig:bumpt-fit}b. The best-fit parameters for the first peak are $F_1 = 0.002 \pm 0.011$~mJy, $t_\mathrm{peak,1} = 76 \pm 6) \times 10^4~\mathrm{s} = 8.8 \pm 0.7$~d
and $\sigma_1 = (3.7 \pm 1.6) \times 10^4~\mathrm{s} = 0.4 \pm 0.2$~d, while the best-fit parameters for the more pronounced second peak are $F_2 = 0.0029 \pm 0.0006$~mJy, $t_\mathrm{peak,2} = (110 \pm 3) \times 10^4~\mathrm{s} = 12.7 \pm 0.3$~d and $\sigma_2 = (12.0 \pm 2) \times 10^4~\mathrm{s} = 1.4 \pm 0.2$~d. The total fluence of the two peaks is $5.73 \times10^{-10}$~\fluence\ and $9.83\times 10^{-10}$~\fluence\ respectively. 

Any re-brightening or flaring episode is likely to have an asymmetric profile with a faster rise and slower decline that is not appropriately modelled by a Gaussian function. Hence, we fit them with a more plausible model, the Norris function \citep{2005ApJ...627..324N}. The intensity of the flare is modelled as

\begin{equation}
I(t) = A \lambda  \exp \left( -\frac{\tau_1}{(t-t_i)} - \frac{(t-t_i)}{\tau_2} \right) \label{eq:norris}
\end{equation}
where $t_i$ is the pulse start time, and the equation holds for $t > t_i$. The parameters $\tau_1$ and $\tau_2$ are associated with the rising and decaying phases of the pulse, but are not directly the rise and decay timescales. The burst intensity is given by the parameter $A\lambda$, where $\lambda = \exp(2\sqrt{\tau_1/\tau_2})$. We ignore the weaker first episode here, but find that the second episode is fit well by Equation~\ref{eq:norris}. The burst ``start time'' is $t_i = 88 \pm 3 \times 10^{4}$. The peak time is $t_\mathrm{peak} = t_i + \sqrt{\tau_1/\tau_2} = (9.9 \pm 2.4) \times 10^5~\mathrm{s} = 11.5 \pm 2.7$~d --- consistent with, but bit sooner than, the values obtained from the double Gaussian fit. The width of the pulse is $w = \tau_2 ( 1 + 4 \sqrt{\tau_1/\tau_2} ) = (3.2 \pm 1.4) \times 10^5~\mathrm{s} = 3.7 \pm 1.6$~d. Under this model, the fluence of the pulse is $1.48\times10^{-9}$~\fluence. For comparison, the total fluence of the underlying afterglow model in the same duration is  $1.85\times 10^{-9}$~\fluence.

In summary, we see evidence for two re-brightening episodes in the afterglow, at about 8.8 and 12.7 days after the burst. The second episode is the more significant, with $\Delta t/t$ values $\sim 0.25 $ and 0.33, and $\Delta F / F \sim$ 1.14 and 1.07 for the double-Gaussian and Norris models, respectively.

\section{Discussion}\label{sec:discussion}
 The afterglow of \thisgrb is quite typical in early times, following a broken power-law behaviour (\S\ref{subsec:brokenpl}) that is modelled well with as a standard afterglow with \af\ (\S\ref{subsec:afterglowmcmc}). The late--time deviations from a smooth decay (\S\ref{subsec:flarefit}) can arise from a variety of reasons in long GRB afterglows. 
 
 A common cause for a re--brightening is the appearance of the supernova associated with the GRB (\S\ref{subsec:sn}). Flaring may also occur due to patchy shells in the jet (\S\ref{subsec:angular}) or interaction of the jet with inhomogeneities the ISM (\S\ref{subsec:ismfluct}). Various shocks can also cause flaring --- for instance a reverse shock in the ejecta (\S\ref{subsec:reverse_shock}) or a collision of two forward shocks (\S\ref{subsec:forward_shock}). Delayed activity by the central engine may manifest directly as flaring (\S\ref{subsec:late_flaring}), or interactions between a delayed jet and a cocoon (\S\ref{subsec:cocoon}), or may refresh the forward shock (\S\ref{subsec:refreshed_shock}). 

We discuss these in detail below, testing each probable cause for the re-brightening in \thisgrb.

\subsection{Supernova}\label{subsec:sn}

\begin{figure}
    \centering
    \includegraphics[width=0.9\columnwidth]{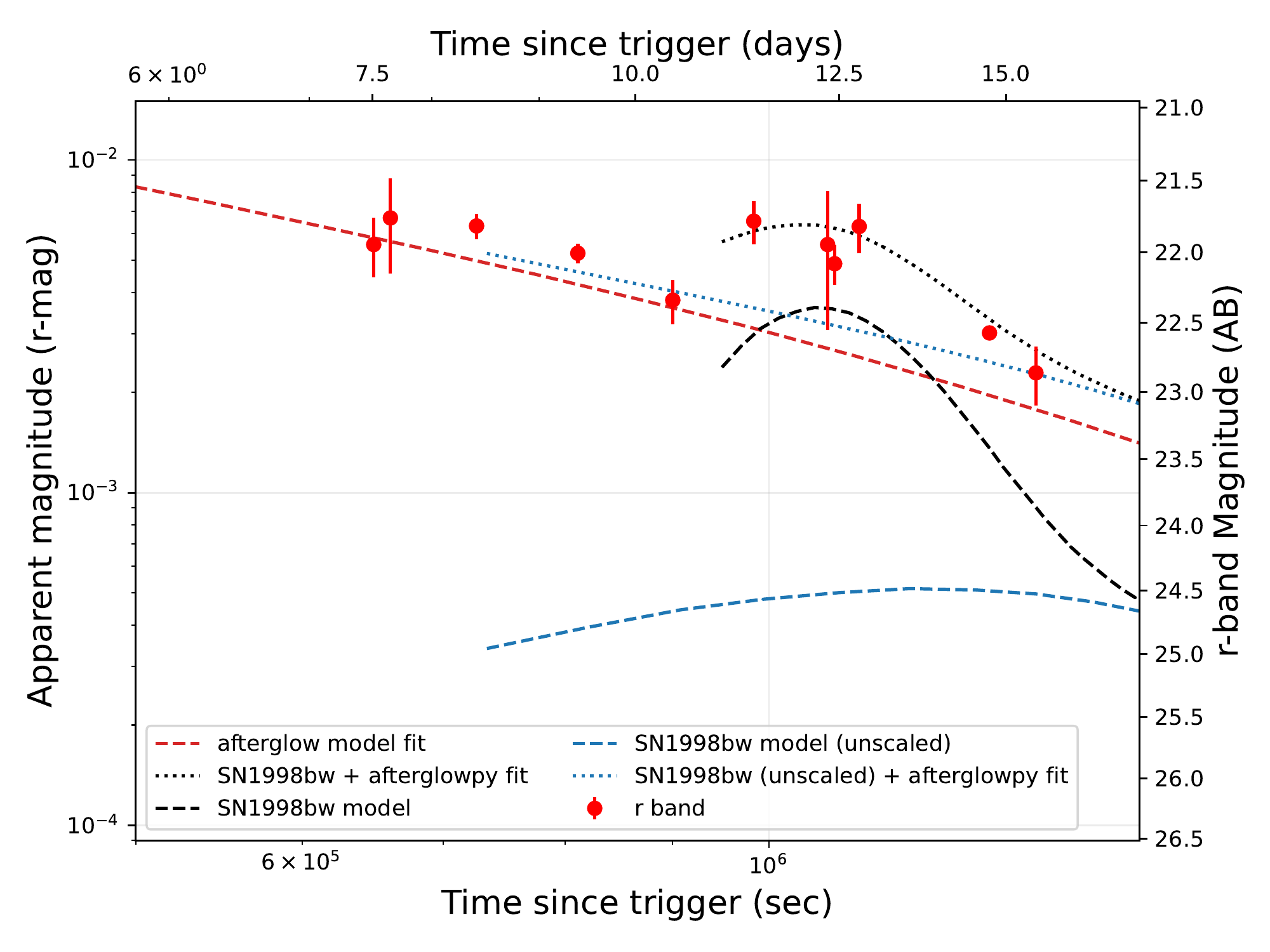}
    \caption{K-corrected SN~1998bw light curve plotted with the excessive emission in \thisgrb. The dashed red line shows the afterglow model, while the dotted black line shows the afterglow and scaled supernova combined model plotted in the observers' frame. The scaled supernova-only model (black dashed line) was obtained by scaling SN~1998bw model (blue dashed line) flux up by 7$\times$, stretched to be eight times faster (shorter) than actual, and shifted in time such that the supernova started six days after the GRB. The unscaled SN~1998bw model fitted over the afterglowpy model is shown with blue dotted lines for comparison. This is an unreasonable set of parameters; thus, a supernova cannot explain the re-brightening.}
    \label{fig:sn_fit}
\end{figure}

 In many long GRBs, the very late-time supernova (SN) bump follows the afterglow emission indicating the collapsar origin for the burst \citep{1998Natur.395..670G, SneGRB, 2021arXiv211011364R}. To test this possibility, we used the light curve of the prototypical SN~1998bw to compare this excessive emission. As a starting point, we referred to SN~1998bw data from \citet{2011AJ....141..163C} and applied a K-correction \citep{2001AJ....121.2879B} using redshift of \thisgrb. Using cubic splines, we interpolated the fluxes into the observed $r$ band values. A continuous light curve was created using cubic interpolation on these values. The resulted light curves were then further scaled in flux (k), stretch in evolution (s) and shifted in time (S$_\mathrm{t}$) to fit \thisgrb light curve. The SN1988bw model light curve overplotted with the \thisgrb data is shown in Figure~\ref{fig:sn_fit}. 
Typical GRB supernovae have absolute magnitudes in the range $-17.5$ to $-20$ with median value $\sim -19.5$ mag \citep{2009AJ....137..347R}, and peak about $\sim 20$ days after the GRB in the rest frame. At the redshift of \thisgrb, it would correspond to an apparent magnitude of 24.6, peaking 38 days after the trigger in the observers' frame. We note that this is the bolometric magnitude, while our observations are in the r band --- corresponding to the rest frame u band. Thus, the expected supernova will be even fainter due to the finite bandwidth and possible extinction. The observed episodes occur much sooner and are much brighter than these values.

Thus, to explain the re-brightening seen in \thisgrb as a supernova similar to SN~1998bw, the SNe light curve has to be made shorter by a factor of 8 ($s = 1/8$), the flux has to be made brighter by a factor of $k \sim$ 7, and the supernova onset has to be delayed by $\sim 6$ days to get a reasonable fit. These parameters --- in particular the shorter timescale and delayed onset of the supernova --- are quite unphysical, and we do not find any acceptable values that can match the light curve. Hence we conclude that the re-brightening is not associated with a supernova.

\subsection{Patchy shell model}\label{subsec:angular}
 The patchy shell model attributes the variability in GRB afterglows to random angular fluctuations in the energy of the relativistic jet \citep{2004ApJ...602L..97N}. However, such variations are expected at earlier times when there are causally disconnected regions within the jet opening angle. However, the variability caused by this mechanism has timescales $\Delta t \gtrsim t$, \citep{2004ApJ...602L..97N,2005ApJ...631..429I}, inconsistent with our measurements. Therefore, we rule out the patchy shell model as a potential cause for the re-brightening seen in \thisgrb.

\subsection{Variations due to fluctuation in ISM density} \label{subsec:ismfluct}

 Ambient density fluctuations can account for late-time variability in GRB afterglows \citep{2000ApJ...535..788W, 2002A&A...396L...5L, 2005ApJ...631..429I}. Such inhomogeneities are primarily caused due to winds from the progenitor or due to turbulence in the ambient medium. \citet{2005ApJ...631..429I} put an upper limit on flux variation due to inhomogeneities in the ambient medium of standard afterglows for on--axis jets:

\begin{equation}
\frac{\Delta F_{\nu}}{F_{\nu}}  \leq  \frac{4}{5} f_{\mathrm{c}}^{-1} \frac{F}{\nu F_{\nu}} \frac{\Delta t}{t}
\end{equation}
Here $f_{\mathrm{c}} \sim ( \nu_{\mathrm{m}}/\nu_{\mathrm{c}})^{(p - 2)/2}$. The light curve of \thisgrb is governed by slow cooling, with optical-band frequencies satisfying  $\nu_{\mathrm{m}} < \nu_{opt}< \nu_{\mathrm{c}}$ criteria at time of excessive emission ($t \sim 12$~days) seen in \thisgrb light curve (see \S\ref{subsec:afterglowmcmc}). For such cases the $F / \nu F_{\nu} \sim (\nu/\nu_{m})^{(p-3)/2}$ \citep{2005ApJ...631..429I}. This suggests $ \frac{\Delta F_{\nu}}{F_{\nu}} \leq 0.14$, which is significantly lower than actual value of 1.07 --- 1.14 (\S\ref{subsec:flarefit}). Further, \citet{2003ApJ...598..400N} show that assuming a spherically symmetric ISM profiles, any flaring from such interactions will have $\Delta t/t > 1$ which is also inconsistent with our measurements. Thus we rule out fluctuations in ambient density as possible origin of flare.

\subsection{Reverse-Shock emission in ejecta medium}\label{subsec:reverse_shock}

 The interaction of blast-wave and circumburst medium results in two shock waves: the forward shock moving towards circumburst medium and a reverse shock moving back into ejecta itself. The reverse shock could produce an optical peak in the observed optical light curve at early times \citep{2003ApJ...582L..75K, 2015ApJ...810..160G, Greiner2009}. Reverse shock is expected to rise rapidly in constant density medium  under thin shell approximation case ($\alpha_{\mathrm{rise}} = 3p - 3/2$, where $p$ is the power-law index of the electron distribution), and decline, with $\alpha_{\mathrm{decline}} = -(27\mathrm{p} + 7)/35$ \citep{2000ApJ...545..807K, Greiner2009}. The canonical range of electron distribution index (p) = $2.2 - 2.5$ \citep{Greiner2009}. The estimated rising and decaying temporal indices of the optical flare are $\sim 6.2$ and $\sim 3.4$, respectively. This implies p $\sim 2.56$ and $4.15$ before and after the peak of the flare, respectively. The inconsistent values of p during the rising and decaying part indicate that the flare is not a result of external reverse shock decay. Moreover, the observed peak time occurs at t$_{\rm peak}$ = (9.90 $\pm$ 2.31) $\times$ 10$^{5}$~s post burst, far later than the $ 10^{3-4}$~s delays expected from flares caused by the reverse shock component in optical bands \citep{2003ApJ...582L..75K,2007ApJ...665L..93U}.

\subsection{Collision of two forward shocks}\label{subsec:forward_shock}

An external collision between shells of GRB can produce flaring on top of afterglow decay \citep{2006ApJ...636L..29P, 2006ApJ...642..354Z, 2005Sci...309.1833B, 2007ApJ...671.1903C}. The time and amplitude and duration of such flares vary among GRBs, depending upon the interaction time, Lorentz factor ($\Gamma$) and the energy $\rm{E}_{iso}$ of the colliding shells. \citet{vlasis} discuss a scenario where a shell with a lower $\Gamma$ is ejected first from the central engine, followed by a shell with a higher $\Gamma$. The first shell decelerates further as it interacts with the interstellar medium, and the second (faster) shell can catch up and ram into the first shell, producing optical flares. For typical GRBs, flares created by such a mechanism should have $\Delta t/ t ~\sim 1$ where $\Delta t$ is the full width at half maximum of the flare, and $t$ is the time at which the flare peaks. We also expect $\Delta F/F ~\sim $ 2-5, where $F$ is the flux of the afterglow and $\Delta F$ is the excess brightening caused by the flare \citep{vlasis}.

The measured $\Delta t/t$ and $\Delta F / F $ values (\S\ref{subsec:flarefit}) are smaller than the predictions of \citet{vlasis}. Thus, we conclude that it is unlikely that collisions between forward shocks caused the late time flaring.

\subsection{Late-time flaring emission from central engine}\label{subsec:late_flaring}

\begin{figure}
    \centering
        \begin{subfigure}{0.8\linewidth}
    \centering
    	\includegraphics[width=\columnwidth]{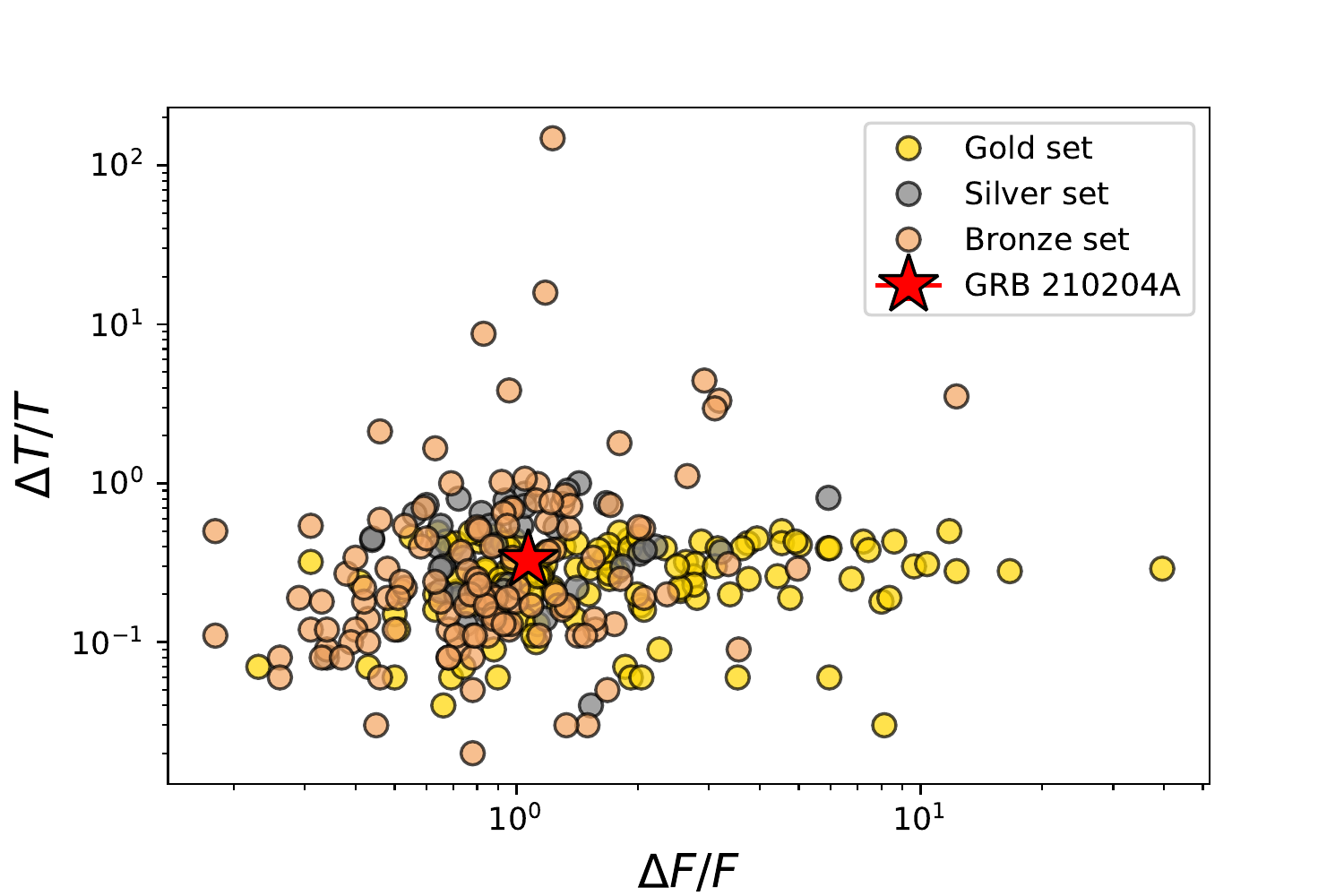}
   	\caption{}
	\label{fig:delt-delf}
    \end{subfigure} 
    \begin{subfigure}{0.8\linewidth}
    \centering
    	\includegraphics[width=\columnwidth]{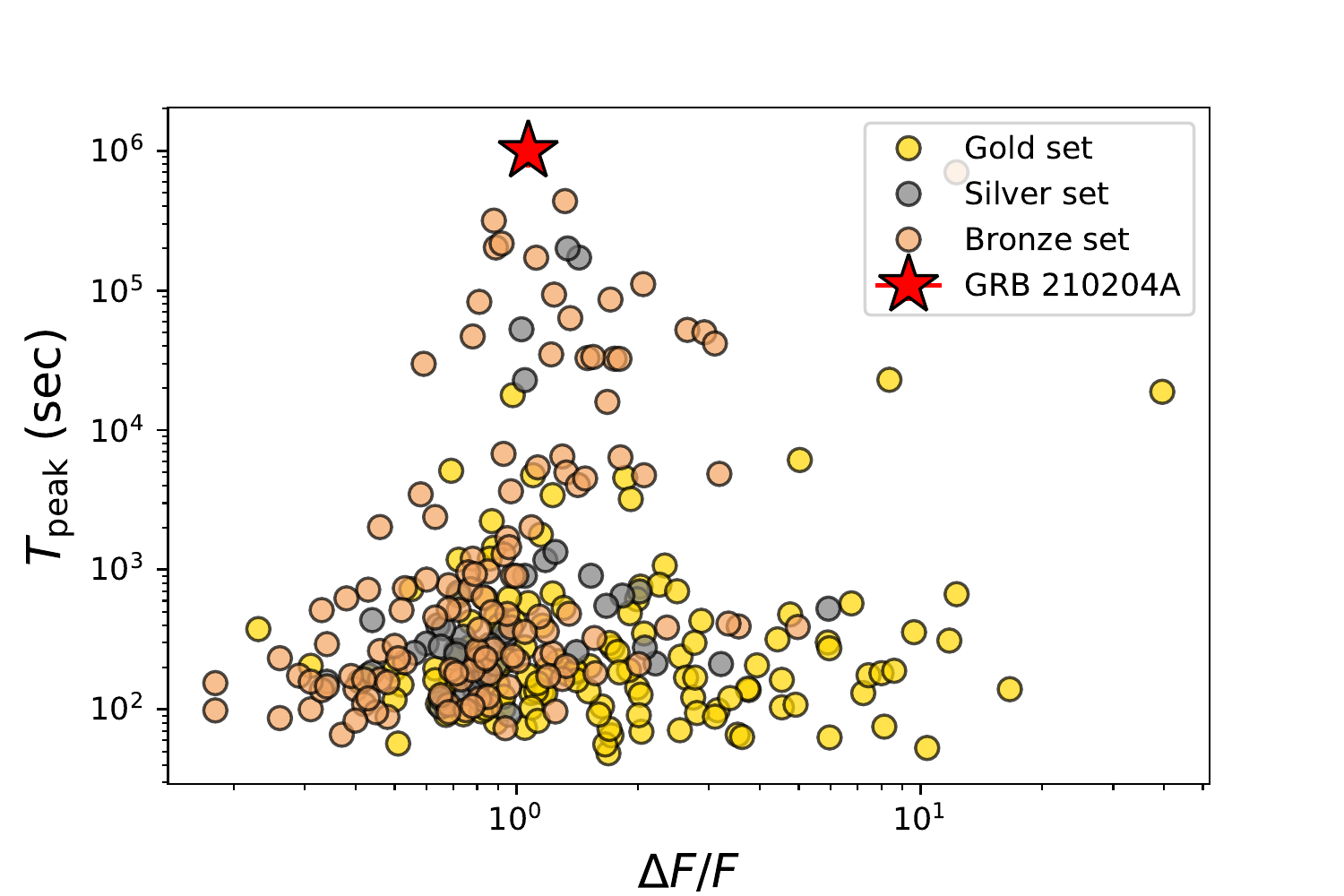}
   	\caption{}
	\label{fig:tpeak-vs-delf}
    \end{subfigure}
    \begin{subfigure}{0.8\linewidth}
    \centering
    	\includegraphics[width=\columnwidth]{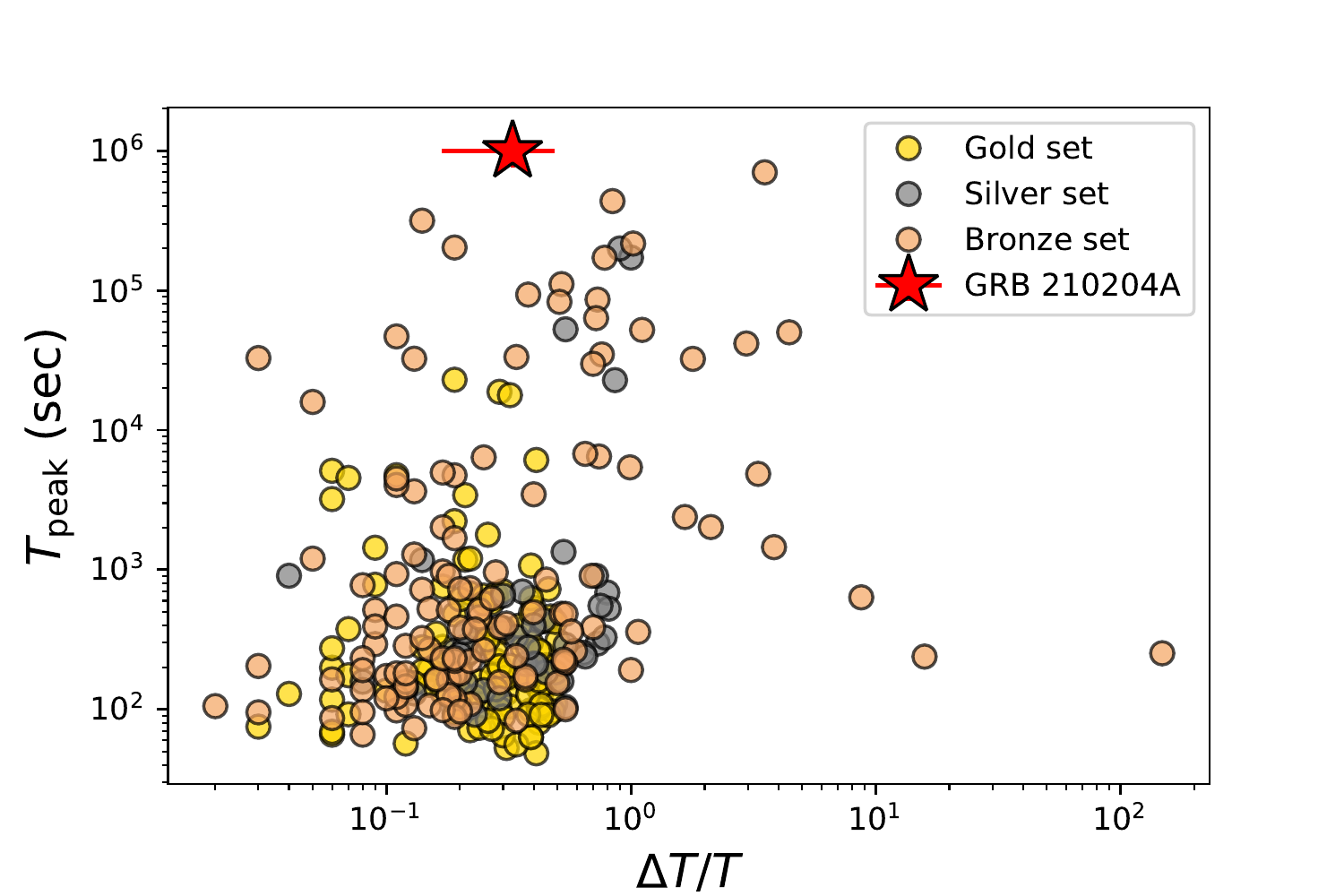}
   	\caption{} 
	\label{fig:tpeak-vs-delt-t}
    \end{subfigure}
     \caption{Comparison of flaring properties of GRB210204A with GRB flares published in \citet{2014ApJ...788...30S}. The gold, silver and bronze sets are shown by respective colors. \thisgrb is depicted with red star symbol, with the limits of the error bar indicating the two $\Delta t$ values measured in \S\ref{subsec:flarefit}. The flare is very similar to other flares in terms of $\Delta F / F$ and $\Delta t/t$, but occurs later than any of the flares in the \citet{2014ApJ...788...30S} data set.}
     \label{fig:flaring_comp}
\end{figure}

 Flaring activity is fairly common in GRB afterglows --- seen in more than 50\% of the GRBs in X-rays \citep{2006ApJ...647.1213O} and $\sim 33\%$ of GRB light curves in optical \citep{Swenson2013}. Due to the very limited amount of X-ray data for \thisgrb, we focus on r-band optical observations here --- in particular, the second re-brightening episode. Flaring in afterglows may be caused by external shocks caused when the jet interacts with density bumps in the interstellar medium (which is discussed in \S\ref{subsec:ismfluct}), or internal shocks from a central engine that is still active, which we discuss here. Indeed, the presence of the three episodes in the prompt emission of \thisgrb is itself an encouraging sign that the central engine is capable of injecting energy multiple times. Such central engine activity itself is typically ascribed to two scenarios. The first possibility is a long-lived magnetar, active to late times \citep{1994AIPC..307..552U,1998A&A...333L..87D, 2000ApJ...545L..73R}. The other possibility is the delayed formation of a black hole in a collapsar, with an accretion disk that may feed matter to the black hole for days \citep{2001ApJ...550..410M}.

Flares are typically characterised by the flaring timescale as compared to the delay time and the fractional increase in flux \citep{2005ApJ...631..429I}. From \S\ref{subsec:flarefit}, we have $\Delta t/t \sim 0.25$ and $\Delta F/F \sim 1.14$. These values are consistent with the classical flaring criteria $\Delta t/t \leq 1$  \citep{Swenson2013,2014ApJ...788...30S}. Next, we compare the properties of the flare with the \citet{Swenson2013} sample. Figure~\ref{fig:delt-delf} shows that the flare is similar to other flares in the duration and flux ratios. What sets it apart is the peak time (Figures~\ref{fig:tpeak-vs-delt-t}). However, this may be an observational bias, as late-time observations by UVOT or other telescopes are not as common. 

Based on the long delay after the GRB, the flares are unlikely to be directly associated with late-time central engine activity. However, we cannot fully rule out this possibility due to the lack of multi-wavelength data.

\subsection{Interaction of a delayed jet with a cocoon}\label{subsec:cocoon}
 The passage of the prompt jet through the stellar envelope creates a cocoon of material \citep{2017ApJ...834...28N}. As the main jet subsides, the cocoon quickly gets filled in due to transverse spreading, presenting a barrier to any delayed jet components like ones discussed in \S\ref{subsec:late_flaring}. \citet{2010MNRAS.403..229S} argue that the interaction of such late jet components with the ejecta can cause broadband flaring. Such flares have $\Delta t/t < 0.5$, and the flux can vary drastically depending on the system parameters. For instance, they predict that for a GRB with redshift 2, the optical $V$-band magnitude for such a flare may be anywhere between 11.5--29, comfortably encompassing the values observed for our $z = 0.876$ case. However, the resultant flares are expected to occur at earlier times: typically starting at 100~s after the burst, but possibly up to 10$^4$--10$^5$~s after the event. The flaring in \thisgrb occurs an order of magnitude later in time, and thus is not any more likely to be caused by interactions between a delayed jet and a cocoon than any other causes of late engine activity discussed in \S\ref{subsec:late_flaring}.

\subsection{Refreshed shock}\label{subsec:refreshed_shock}

The late-time brightening in the optical light curve could also have originated from a forward shock that is refreshed by late-time energy injection from the central engine \citep{1998ApJ...496L...1R, 1998ApJ...503..314P}. Such a refreshed shock scenario has been used to explain the observed re-brightening in the optical light curves of GRB~030329 \citep{2003Natur.426..138G} and GRB~120326A \citep{2014A&A...572A..55M}. Consider a standard forward shock model where the bulk Lorentz factor of the ejecta is not constant but has a range of values. Faster moving shells with higher Lorentz factors ($\Gamma_{\rm fast} \geq$ 100) interact with the surrounding medium first and are slowed down. Slower moving shells ($\Gamma_{\rm slow} \sim$ 10) catch up with these decelerated shells at late times, injecting energy into the shock and increasing the emission. \citet{2006AIPC..836..353G} derived the formula for the collision time of two shells (one moving with $\Gamma_{\rm slow} \sim$ 10 and other moving with $\Gamma_{\rm fast} \geq$ 100) considering the simple assumption (see equation \ref{Gamma1}), 

\vspace{-0.5cm}
\begin{eqnarray}
t_{\rm shock} \approx 1.66 \times E_{\rm \gamma,iso,53}^{1/3}~n_{0}^{-1/3}~\Gamma_{\rm slow, 10}^{-8/3} ~~~~{\rm days}
\label{Gamma1}
\end{eqnarray}

In this equation, $E_{\rm \gamma,iso,53}$ denotes $E_{\rm \gamma,iso}$/10$^{53}$ in erg, n$_{0}$ is the density for a constant medium which we obtained from broadband afterglow modelling, $\Gamma_{\rm slow, 10}$ is the bulk Lorentz factor of slow moving shell in the unit of 10. Following the above equation, we calculated the Lorentz factor of the slow moving shell for \thisgrb at the time of optical brightening. We take $t_{\rm shock} \sim 12.7$~days from our two-Gaussian fit (\S\ref{subsec:flarefit}), and substitute $E_{\rm \gamma,iso,53} = 10^{1.06}$~erg, n$_{0} = 10^{-5.65}~\mathrm{cm}^{-3}$ from Table~\ref{tab:mcmc_post} to get $\Gamma_{\rm slow} \sim 32$. 

For flares caused by refreshed shocks, we expect $\Delta t /t > 1/4$, broadly consistent with the values measured in \S\ref{subsec:flarefit}. Thus, a refreshed shock scenario is a plausible explanation for these flares.

\section{Summary and Conclusion}

We presented a detailed analysis of the prompt emission and afterglow of \thisgrb. The prompt emission consists of three distinct emission episodes in \fermi-GBM data, separated by quiescent phases. Spectral analysis of the third and brightest episode shows the presence of a thermal component at low energies, adding a member to the small but growing class of GRBs with thermal components. We also find that \thisgrb (full interval), as well as the individual pulses, are consistent with the Amati relation.

\thisgrb stands out by having the most delayed flaring activity ever detected in GRBs. A flare is detected 8.8~days after the burst, followed by a stronger flare at 12.7~days. We analyse a multitude of possible causes for such flaring and rule out most of them. We conclude that the flaring is likely caused by late-time activity in the central engine --- manifesting either as flares caused due to internal shocks, the interaction of a delayed jet with a cocoon or by refreshing a forward shock. 

Such late-time data are not available for most GRBs. It is plausible that more GRBs exhibit such late time flaring activity, but the sample suffers severally from observational biases. This underscores the need for a systematic follow-up program for GRB afterglows. We have undertaken such a program with the GROWTH-India telescope to probe afterglow features in detail.

\section*{Acknowledgements}

This work made use of data from the \git~(GIT) set up by the Indian Institute of Astrophysics (IIA) and the Indian Institute of Technology Bombay (IITB). It is located at the Indian Astronomical Observatory (Hanle), operated by IIA. We acknowledge funding by the IITB alumni batch of 1994, which partially supports operations of the telescope. Telescope technical details are available at \url{https://sites.google.com/view/growthindia/}.

This work is partially based on data obtained with the 2m Himalayan Chandra Telescope of the Indian Astronomical Observatory (IAO), operated by the Indian Institute of Astrophysics (IIA), an autonomous Institute under Department of Science and Technology, Government of India. We thank the staff at IAO and at IIA's Centre for Research and Education in Science and Technology (CREST) for their support. 

We thank Jesper Sollerman for his useful suggestions that helped in improving quality of this work. 

This research is partially based on observations (proposal number DOT-2021-C1-P62; PI: Rahul Gupta and DOT-2021C1-P19; PI: Ankur Ghosh) obtained at the 3.6m Devasthal Optical Telescope (DOT), which is a National Facility run and managed by Aryabhatta Research Institute of Observational Sciences (ARIES), an autonomous Institute under Department of Science and Technology, Government of India.

PC acknowledges support of the Department of Atomic Energy, Government of India, under project no. 12-R\&D-TFR-5.02-0700. We thank the staff of the GMRT that made these observations possible. The GMRT is run by the National Centre for Radio Astrophysics of the Tata Institute of Fundamental Research.

Harsh Kumar thanks the LSSTC Data Science Fellowship Program, which is funded by LSSTC, NSF Cybertraining Grant \#1829740, the Brinson Foundation, and the Moore Foundation; his participation in the program has benefited this work.

RG, AA, VB, KM, and SBP acknowledge BRICS grant {DST/IMRCD/BRICS /PilotCall1 /ProFCheap/2017(G)} for the financial support. RG, VB, and SBP also acknowledge the financial support of ISRO under AstroSat archival Data utilization program (DS$\_$2B-13013(2)/1/2021-Sec.2). RG is also thankful to Dr P. Veres for sharing data files presented in Figure \ref{fig:T90HR}. This publication uses data from the AstroSat mission of the Indian Space Research Organisation (ISRO), archived at the Indian Space Science Data Centre (ISSDC).

 This research has made use of data obtained from \hct~under proposal number HCT-2021-C1-P02. We thank HCT stuff for undertaking the observations. HCT observations were carried out under the ToO program.

This research has made use of the NASA/IPAC Extragalactic Database (NED), which is funded by the National Aeronautics and Space Administration and operated by the California Institute of Technology. 

This research has made use of data obtained from the High Energy Astrophysics Science Archive Research Center (HEASARC) and the Leicester Database and Archive Service (LEDAS), provided by NASA's Goddard Space Flight Center and the Department of Physics and Astronomy, Leicester University, UK, respectively.

This research has made use of NASA's Astrophysics Data System.

This research has made use of data and/or services provided by the International Astronomical Union's Minor Planet Center. 

This research has made use of the VizieR catalogue access tool, CDS, Strasbourg, France (DOI : 10.26093/cds/vizier). The original description of the VizieR service was published in 2000, A\&AS 143, 23.


\section*{Data Availability}
All data used in this article have been included in a tabular format within the article.

\bibliographystyle{mnras}
\bibliography{ref.bib} 

\appendix
\section{Additional figures from afterglow fits}
\onecolumn
 \begin{figure*}
    \centering
    \includegraphics[width=0.95\textwidth]{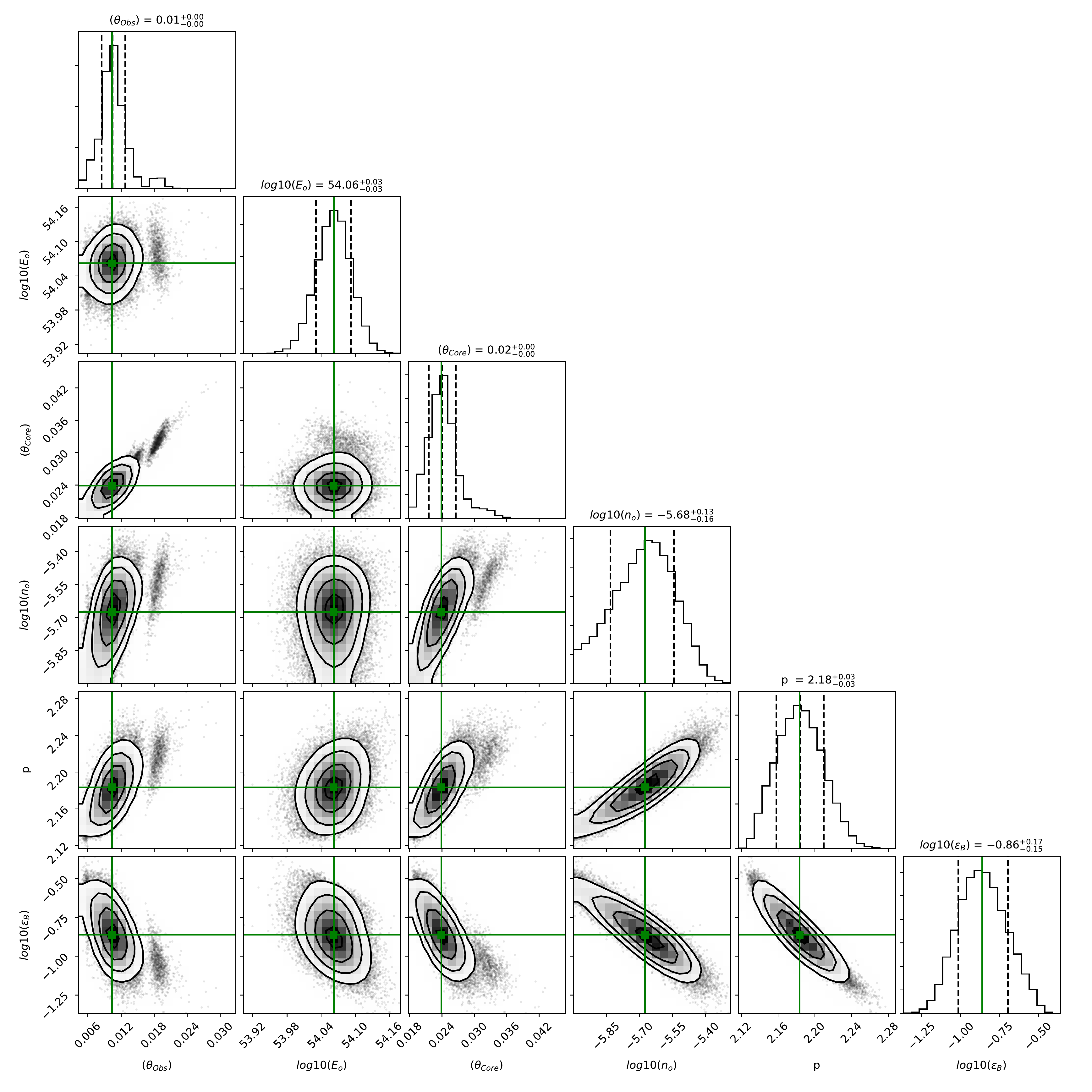}
    \caption{Posterior distribution of parameters for model fitted using $\af$ and \sw{EMCEE}. The model fit for the $\rm{\theta}_{obs}$, $\rm{\theta}_{core}$, $\rm{log_{10}}(n_{0})$, p, $\rm{log_{10}}(E_{0})$,  $\rm{log_{10}}(\epsilon_{B})$ parameters. the histogram shows the 16\%, 50\% and 84\% percentiles of probability distribution.}
    \label{fig:corner}
\end{figure*}

\clearpage

\begin{longtable}{ccccc}
\caption{Log of our photometry observations of the optical afterglow of \thisgrb, taken with various ground-based telescopes.} \label{table:phot-table-our} \\
\hline
\textbf{JD} & \textbf{T-T0 (sec)} & \textbf{Filter} & \textbf{Magnitude} &  \textbf{Telescope/Instrument} \\
\hline
2459248.74875 & -88273.15200 & r & > 20.87  & P48$+$ZTF \\
2459248.76859 & -86558.15520 & g & > 21.34  & P48$+$ZTF \\
2459248.83259 & -81029.15136 & i & > 20.22  & P48$+$ZTF \\
2459249.71868 & -4471.14816 & g & > 18.74  & P48$+$ZTF \\
2459249.79662 & 2262.85056 & r & 17.16 $\pm$ 0.03 & P48$+$ZTF \\
2459250.71466 & 81581.85216 & r & 19.31 $\pm$ 0.06 & P48$+$ZTF \\
2459250.75523 & 85086.84960 & g & 19.75 $\pm$ 0.08 & P48$+$ZTF \\
2459251.72700 & 169047.84672 & r & 19.99 $\pm$ 0.08 & P48$+$ZTF\\
2459251.73331 & 169592.84928 & i & 19.71 $\pm$ 0.10 & P48$+$ZTF\\
2459251.89443 & 183513.85056 & g & 20.62 $\pm$ 0.2 & P48$+$ZTF\\
2459253.72924 & 342041.84928 & i & > 19.80 & P48$+$ZTF \\
2459254.82180 & 436438.85184 & i & > 20.70  & P48$+$ZTF \\
2459257.80761 & 694412.84448 & i & > 20.00 & P48 +  ZTF \\
2459260.82290 & 954933.84864 & i & > 19.70 & P48$+$ZTF \\ \hline
2459252.35734 & 223509.45600 & r & 20.55 $\pm$ 0.03 & GIT \\
2459253.20464 & 296715.74400 & r & 20.89 $\pm$ 0.03 & GIT \\
2459254.29448 & 390877.92000 & r & 21.31 $\pm$ 0.04 & GIT \\
2459255.32039 & 479516.97600 & r & 21.72 $\pm$ 0.05 & GIT \\
2459266.24093 & 1423051.63200 & r & > 22.39  & GIT \\
2459267.194 & 1505396.44800 & r & > 22.30 & GIT \\
2459271.19364 & 1850965.34400 & r & > 20.81  & GIT \\
2459272.22723 & 1940267.52000 & r & > 20.57  & GIT \\
2459274.17682 & 2108712.09600 & r & > 21.69  & GIT \\
2459275.17860 & 2195266.32000 & r & > 22.0 & GIT \\ \hline
2459252.13277 & 204106.17600 & R & 20.17 $\pm$ 0.05 &HCT \\
2459252.13753 & 204517.44000 & R & 20.16 $\pm$ 0.05 &HCT \\
2459252.14045 & 204769.72800 & R & 20.15 $\pm$ 0.03 &HCT \\
2459252.14231 & 204930.43200 & R & 20.20 $\pm$ 0.06 &HCT \\
2459252.15787 & 206274.81600 & I & 19.77 $\pm$ 0.06 &HCT \\
2459252.16027 & 206482.17600 & I & 19.69 $\pm$ 0.04 &HCT \\
2459252.16265 & 206687.80800 & I & 19.74 $\pm$ 0.06 &HCT \\
2459252.16742 & 207099.93600 & I & 19.78 $\pm$ 0.06 &HCT \\
2459253.2722 & 302552.92800 & R & 20.67 $\pm$ 0.05 &HCT \\
2459253.28418 & 303588.00000 & I & 20.17 $\pm$ 0.04 &HCT \\
2459258.08132 & 718060.89600 & I & 21.30 $\pm$ 0.20 &HCT \\
2459258.10539 & 720140.54400 & V & 22.14 $\pm$ 0.11 &HCT \\
2459260.18751 & 900035.71200 & R & 22.22 $\pm$ 0.15 & HCT \\ \hline
2459252.18709 & 208799.84822 & R & 20.19 $\pm$ 0.03 & DFOT \\
2459259.15934 & 811201.82400 & R & 21.86 $\pm$ 0.07 & DFOT \\
2459259.34474 & 827220.38400 & I & 21.60 $\pm$ 0.10 & DFOT \\\hline
2459252.18877 & 208944.96221 & R & 20.22 $\pm$ 0.04 & DOT \\
2459252.19235 & 209255.02675 & R & 20.24 $\pm$ 0.04 & DOT  \\
2459252.19528 & 209508.05174 & I & 19.64 $\pm$ 0.05 & DOT  \\
2459252.19748 & 209698.10842 & I & 19.68 $\pm$ 0.05 & DOT  \\
2459252.20512 & 210359.26022 & V & 20.62 $\pm$ 0.05 & DOT  \\
2459252.20909 & 210701.33424 & B & 21.18 $\pm$ 0.09 & DOT  \\
2459252.21269 & 211011.89040 & B & 21.30 $\pm$ 0.06 & DOT  \\
2459252.21269 & 211011.89213 & B & 21.18 $\pm$ 0.08 & DOT \\
2459252.21669 & 211357.47053 & B & 21.13 $\pm$ 0.09 & DOT  \\
2459252.26456 & 215493.34522 & I & 19.67 $\pm$ 0.04 & DOT  \\
2459252.26845 & 215829.42480 & R & 20.29 $\pm$ 0.05 & DOT  \\
2459252.27239 & 216169.98163 & V & 20.69 $\pm$ 0.06 & DOT  \\
2459252.27631 & 216509.06707 & B & 21.13 $\pm$ 0.08 & DOT  \\
2459252.27990 & 216819.10569 & B & 21.20 $\pm$ 0.09 & DOT \\
2459252.28421 & 217191.69187 & I & 19.66 $\pm$ 0.04 & DOT \\
2459252.28808 & 217525.78684 & R & 20.37 $\pm$ 0.06 & DOT \\
2459252.29246 & 217903.83609 & V & 20.69 $\pm$ 0.06 & DOT \\
2459252.35158 & 223012.44518 & I & 19.81 $\pm$ 0.07 & DOT \\
2459252.35554 & 223354.01980 & R & 20.34 $\pm$ 0.07 & DOT \\
2459252.35939 & 223687.06848 & V & 20.72 $\pm$ 0.08 & DOT \\
2459252.36324 & 224019.63072 & B & 21.29 $\pm$ 0.12 & DOT \\
2459252.36688 & 224333.71804 & B & 21.31 $\pm$ 0.12 & DOT \\
2459252.37481 & 225019.35734 & I & 19.78 $\pm$ 0.07 & DOT \\
2459252.37847 & 225335.42669 & I & 19.81 $\pm$ 0.07 & DOT \\
2459252.38265 & 225696.52339 & R & 20.37 $\pm$ 0.08 & DOT \\
2459253.12825 & 290116.63123 & R & 20.76 $\pm$ 0.05 & DOT \\
2459253.13132 & 290381.68829 & I & 20.20 $\pm$ 0.05 & DOT \\
2459253.13448 & 290654.74858 & V & 21.09 $\pm$ 0.05 & DOT \\
2459253.13721 & 290890.30608 & B & 21.59 $\pm$ 0.09 & DOT \\
2459253.13964 & 291100.84646 & B & 21.57 $\pm$ 0.09 & DOT \\
2459253.17108 & 293816.41402 & I & 20.25 $\pm$ 0.05 & DOT \\
2459253.17355 & 294030.45274 & R & 20.76 $\pm$ 0.05 & DOT \\
2459253.17646 & 294281.51472 & V & 21.11 $\pm$ 0.05 & DOT \\
2459253.23432 & 299280.58329 & I & 20.22 $\pm$ 0.05 & DOT \\
2459253.23678 & 299493.63014 & R & 20.79 $\pm$ 0.06 & DOT \\
2459253.23947 & 299725.66166 & R & 20.79 $\pm$ 0.12 & DOT \\
2459253.24264 & 299999.70950 & B & 21.58 $\pm$ 0.08 & DOT \\
2459253.28455 & 303620.97456 & I & 20.33 $\pm$ 0.11 & DOT \\
2459253.28769 & 303892.56432 & R & 20.78 $\pm$ 0.12 & DOT \\
2459253.29160 & 304230.13776 & B & 21.70 $\pm$ 0.09 & DOT \\
2459253.35684 & 309866.34932 & I & 20.34 $\pm$ 0.07 & DOT \\
2459253.35955 & 310100.86915 & R & 20.88 $\pm$ 0.06 & DOT \\
2459253.36224 & 310332.94905 & R & 20.87 $\pm$ 0.13 & DOT \\
2459262.54515 & 1103736.61238 & r & 21.90 $\pm$ 0.17 & DOT \\
2459264.50267 & 1272865.95763 & r & 22.70 $\pm$ 0.05 & DOT \\
2459265.27083 & 1339234.84771 & r & 23.00 $\pm$ 0.20 & DOT \\
2459265.43097 & 1353071.30659 & r & > 22.59  & DOT \\
2459266.45952 & 1441938.10176 & r & > 21.44 & DOT \\
2459267.38564 & 1521954.81360 & r & > 21.13  & DOT \\
2459269.53000 & 1707227.09337 & r & > 19.14  & DOT \\
\hline
\end{longtable}

\begin{longtable}{cccccc}
\caption{Photometry table of the optical afterglow of \thisgrb, data obtained from various reported GCNs.}
\label{tab:public}\\
\hline
\textbf{JD} & \textbf{T-T0 (sec)} & \textbf{Filter} & \textbf{Magnitude} &  \textbf{Instrument} & \textbf{Reference} \\
\hline
2459252.0362 & 195764 & R & 19.94 $\pm$ 0.09 & AZT-33IK & 29417  \\
2459252.2179 & 211462 & R & 20.1 $\pm$ 0.04 & AS-32 & 29417 \\
2459252.8472 & 265835 & g & 21.10 $\pm$ 0.10 & LBT & 29433 \\
2459252.8472 & 265835 & r & 20.70 $\pm$ 0.10 & LBT & 29433 \\
2459252.8472 & 265835 & i & 20.40 $\pm$ 0.10 & LBT  & 29433 \\
2459252.8472 & 265835 & z & 20.2  $\pm$ 0.10 & LBT	& 29433 \\
2459253.0930 & 287073 & R & 20.61 $\pm$ 0.04 & AZT-33IK & 29438 \\
2459254.1607 & 379326 & R & 20.92 $\pm$ 0.05 & AZT-33IK & 29438 \\
2459255.2961 & 477422 & R & 21.09 $\pm$ 0.08 & DFOT & 29490 \\
2459255.3265 & 480047 & R & 21.4 $\pm$ 0.20 & ZTSh & 29499 \\
2459257.2800 & 648835 & R & 21.8 $\pm$ 0.20 & AS-32 & 29499 \\
2459257.4186 & 660802 & R & 21.6 $\pm$ 0.30 & AS-32 & 29499 \\
2459258.1744 & 726104 & R & 21.66 $\pm$ 0.09 & AZT-33IK & 29520 \\
2459261.1528 & 983436 & r & 21.86 $\pm$ 0.15 & AZT-20 & 29520 \\
2459262.1110 & 1066228 & R & 21.8 $\pm$ 0.40 & AZT-33IK & 29520 \\
2459262.2063 & 1074461 & r & 22.18 $\pm$ 0.14 & AZT-20 & 29520 \\
\hline
\end{longtable}

%

\begin{longtable}{cccc}
\caption{Log of X-ray observations of the X-ray afterglow of \thisgrb taken using \swift XRT in 10~\keV band. This data uses a absorption of  $0.61 \times 10^{22} \rm{cm}^{-2}$ at $z$ = 0.876.}
\label{tab:xrt} \\
\hline
\textbf{JD} & \textbf{T-T0 (sec)} & \textbf{Photon Index} & \textbf{Flux Density ($\mu$Jy)} \\
\hline
2459251.634  & 161000.965 & $1.61^{+0.29}_{-0.22} $ & 140.99 $\pm$ 31.98 \\
2459251.636  & 161214.462 & $1.62^{+0.28}_{-0.21} $ & 124.61 $\pm$ 28.42 \\
2459251.640  & 161501.848 & $1.65^{+0.27}_{-0.20} $ & 138.95 $\pm$ 24.97 \\
2459251.702  & 166845.863 & $ 2.11^{+0.24}_{-0.19} $ & 42.26 $\pm$ 9.79 \\
2459251.704  & 167087.251 & $2.13^{+0.25}_{-0.20} $ & 35.41 $\pm$ 9.41 \\
2459251.706  & 167273.799 & $2.15^{+0.25}_{-0.20} $ & 51.94 $\pm$ 11.80 \\
2459251.708  & 167449.688 & $2.14^{+0.25}_{-0.20} $ & 59.20 $\pm$ 13.42 \\
2459251.712  & 167726.371 & $2.11^{+0.24}_{-0.19} $ & 32.05 $\pm$ 8.73 \\
2459251.767  & 172498.443 & $1.64^{+0.29}_{-0.22} $ & 118.14 $\pm$ 26.65 \\
2459251.769  & 172708.123 & $1.62^{+0.30}_{-0.23} $ & 107.96 $\pm$ 27.21 \\
2459251.773  & 173048.102 & $1.61^{+0.30}_{-0.23} $ & 80.00 $\pm$ 17.04 \\ 
2459253.174  & 294064.238 & $1.46^{+0.30}_{-0.22} $ & 63.53 $\pm$ 16.86 \\ 
2459253.178  & 294390.208 & $1.46^{+0.30}_{-0.22} $ & 96.35 $\pm$ 25.20 \\
2459253.182  & 294775.487 & $1.48^{+0.29}_{-0.22} $ & 66.52 $\pm$ 17.42 \\
2459253.187  & 295152.375 & $1.52^{+0.27}_{-0.20} $ & 61.49 $\pm$ 15.55 \\
2459253.239  & 299664.401 & $2.0^{+0.29}_{-0.21} $ & 35.88 $\pm$ 9.36 \\
2459253.243  & 300009.566 & $2.04^{+0.30}_{-0.23} $ & 25.27 $\pm$ 6.62 \\
2459253.247  & 300337.201 & $2.08^{+0.32}_{-0.24} $ & 42.59 $\pm$ 11.11 \\
2459253.250  & 300677.495 & $2.11^{+0.34}_{-0.26} $ & 27.84 $\pm$ 7.04 \\
\hline
\end{longtable}

\begin{longtable}{cccc}
\caption{Log of radio data for the radio afterglow of \thisgrb taken using uGMRT.} 
\label{tab:radio}\\
\textbf{JD} & \textbf{T-T0 (sec)} & \textbf{Energy-Band} & \textbf{Flux ($\mu$Jy)}\\
\hline
2459266.06 	& 1402272.00  & 1254.6 \GHz  & 140 $\pm$ 22 \\
2459281.09  & 2706011.71  & 1254.6 \GHz  & 130 $\pm$ 20\\
2459283.06  & 2876272.416 & 647.8 \GHz   & 95 $\pm$  45 \\
\hline
\end{longtable}


\bsp	
\label{lastpage}
\end{document}